\documentclass[aps,prb,reprint,superscriptaddress,amsmath,nofootinbib]{revtex4-1}
\usepackage{bm,amsmath}
\usepackage{ifpdf}
\usepackage{siunitx}
\usepackage{braket}
\usepackage{comment}

\usepackage[whole]{bxcjkjatype} 
\usepackage{graphicx}
\usepackage[colorlinks=true,urlcolor=blue,citecolor=blue,linkcolor=blue,breaklinks=true]{hyperref}
\usepackage{txfonts}
\usepackage{xcolor}

\graphicspath{{figs/}}

\DeclareGraphicsExtensions{.pdf}

\begin{document}

\title{Floquet topological $d+id$ superconductivity induced by chiral many-body interactions}

\author{Sota~Kitamura}
\affiliation{Department of Applied Physics, The University of Tokyo, Hongo, Tokyo, 113-8656, Japan}

\author{Hideo~Aoki}
\affiliation{Department of Physics, University of Tokyo, Hongo, Tokyo 113-0033, Japan}
\affiliation{Electronics and Photonics Research Institute, Advanced Industrial Science and Technology (AIST), Tsukuba, Ibaraki 305-8568, Japan}

\begin{abstract}
We study how a $d$-wave superconductivity is changed when 
illuminated by circularly-polarised light (CPL) in the repulsive 
Hubbard model in the strong-coupling regime.  
We adopt the Floquet formalism for the Gutzwiller-projected 
effective Hamiltonian with the time-periodic Schrieffer-Wolff transformation.  
We find that CPL induces a topological superconductivity 
with a $d+id$ pairing, which arises from the 
chiral spin coupling and the three-site term 
generated by the CPL.  The latter term remains significant 
even for low frequencies and low intensities of the CPL.  
This is clearly reflected in the obtained phase diagram against the laser intensity 
and temperature for various frequencies red-detuned from 
the Hubbard $U$, with the transient dynamics also examined.  
The phenomenon revealed here can open a novel, dynamical 
way to induce 
a topological superconductivity.  
\end{abstract}

\date{\today}

\maketitle

\section{Introduction}

\begin{figure*}[t]
\begin{centering}
\includegraphics[width=.8\linewidth]{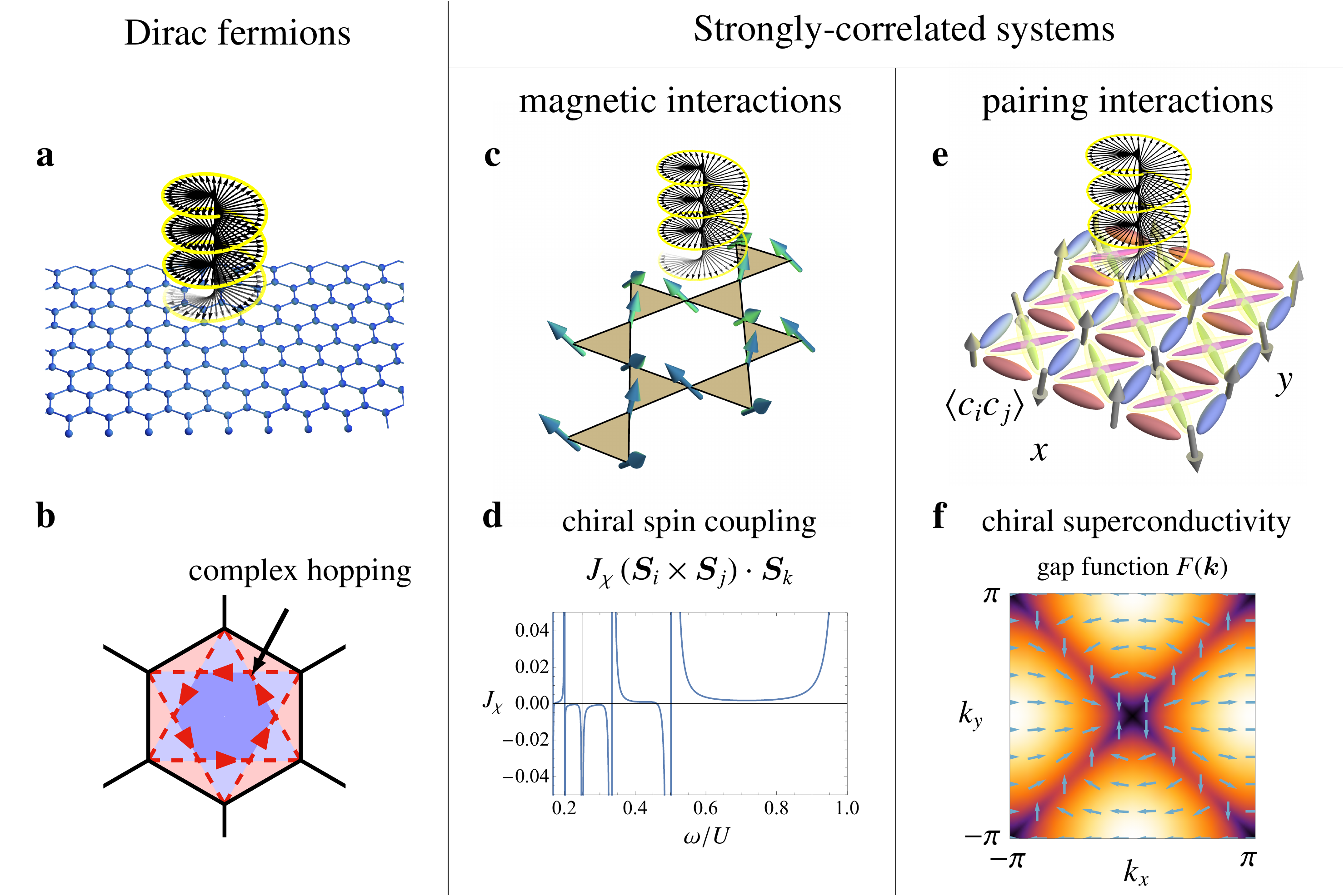}
\par\end{centering}
\caption{\textbf{Various Floquet-induced quantum states.}  
For Dirac fermions as in graphene, 
\textbf{a}  
illumination of a circularly-polarised light (CPL) 
induces the Floquet topological insulator, 
which arises from photon-assisted complex 
hopping (\textbf{b}), where colour shadings 
represent the piece-wise magnetic fluxes. 
\textbf{c} For magnetic phases in strongly-correlated 
systems, 
illumination of CPL induces 
a chiral spin coupling, for which \textbf{d} 
illustrates a resonant behaviour at 
$\omega/U = 1, 1/2, 1/3, \cdots$.  
\textbf{e} For superconductors (present work), illumination of CPL 
on a $d_{x^2-y^2}$-wave superconductor produces 
pairing amplitudes $\langle\hat{c}_{i\uparrow}\hat{c}_{j\downarrow}\rangle$ across nearest neighbours (red: positive; 
blue: negative) along with 
imaginary diagonal bonds (magenta and green), 
leading to an emergent complexified 
gap function $F(\bm{k})$, 
and we have a photo-induced chiral $d_{x^2-y^2}+id_{xy}$ superconductivity (\textbf{f}), where arrows schematically indicate the phase of the complex gap function.}
\label{fig:overview}
\end{figure*}

It has become one of the key pursuits in the condensed-matter physics to design new quantum phases 
as exemplified by unconventional superconductivity 
and topological states.  
Conventionally, materials 
design is the accepted way, in which 
we tailor the crystal structures and consituent 
elements, as combined, if necessary, with 
carrier doping, pressure applications, etc.  
An entirely different avenue should be a 
``non-equilibrium design", in which we envisage 
to realise interesting quantum phases by putting 
the systems out of equilibrium, typically by 
illuminating intense laser lights.  
This opens an in-situ 
way to convert the system that would be unthinkable 
in equilibrium, and is also of 
fundamental interests as non-equilibrium physics.  
Ilya Prigogine, in his book \textit{From being to becoming}~\cite{Prigogine1980},
once said that he would have liked to entitle the 
book as \textit{Time, the forgotten dimension}, 
and we can indeed enlarge our horizon if we do not 
forget the temporal dimension.  

One of the most important pathways is the Floquet physics, 
with which 
we can seek various novel quantum states arising 
from application of AC modulations to the system.  
This is based on Floquet's theorem for time-periodic modulations, put forward by Gaston Floquet in 1883, which 
is much older than 1928 theorem by Bloch for spatially-periodic 
modulations.  A prime application of Floquet physics is 
the 
``Floquet topological insulator" proposed by Takashi 
Oka and one of the present authors in 2009~\cite{Oka2009,Oka2010,Oka2011}.
Namely, by applying a circularly-polarised light to 
honeycomb systems such as graphene [see Fig.~\ref{fig:overview}a], we can turn the 
system into a topological insulator in a dynamical manner.  In other words, we are here talking about 
matter-light combined states, since the electron is converted into a superposition 
of the original, one-photon dressed, two-photon dressed, 
..., electrons in the Floquet picture.  
The Floquet topological insulator with a topological gap 
exhibits a DC Hall effect despite the modulation being 
AC.  
The resulting state shows a kind of quantum anomalous 
Hall effect (i.e., quantum Hall effect in zero 
magnetic field) originally proposed for the static case 
by Duncan Haldane back 
in 1988~\cite{Haldane1988}.  Indeed, the effective Hamiltonian 
for the irradiated system in the leading order 
in the Floquet formalism exactly coincides with 
Haldane's model [Fig.~\ref{fig:overview}b], as shown by Takuya Kitagawa \textit{et al.}~\cite{Kitagawa2011}  
The year 2019 witnessed an 
experimental detection of the Floquet topological 
insulator in graphene by James McIver \textit{et al.}~\cite{McIver2019}

Thus we have a surge of interests in 
Floquet topological phases, for many-body problems 
as well as one-body cases.  Floquet physics has also 
been extended to explore superconductivity in 
AC-modulated situations.  It is out of the scope of 
the present paper to review the whole field, but 
let us briefly mention that the spectrum of 
the many-body interests in Floquet physics 
covers a range of quantum phases as 

\begin{table}[h]
\begin{tabular}{c|c}
topology & superconductivity \\
\hline 
FTI $\leftrightarrow$ Mott's insulator & Attraction-repulsion conversion \\
chiral spin states & $\eta$-pairing
\end{tabular}
\end{table}
 
\noindent First, when a repulsive Hubbard model 
for correlated electrons is illuminated by laser, 
phase transitions arise between 
the Floquet topological insulator (FTI) and Mott's 
insulator on a phase diagram 
against the AC modulation intensity and the 
repulsive electron-electron interaction.\cite{Mikami2016} 
Second, circularly-polarised 
laser lights can induce chiral spin interactions, 
$(\hat{\bm{S}}_{i}\times\hat{\bm{S}}_{j})\cdot\hat{\bm{S}}_{k}$, and 
consequent spin liquids for 
the strongly-coupled systems with the repulsion $U$ much greater than the electron hopping $t_0$ 
[see Fig.~\ref{fig:overview}c.\ref{fig:overview}d]~\cite{Takayoshi2014,Takayoshi2014-2,Kitamura2017,Claassen2017}. 
There, we can realise that the Floquet-incuded chiral coupling becomes significant 
when the frequency $\omega$ of 
the laser is set close to the Hubbard $U$, 
with vastly different behaviours 
between the cases when $\omega$ 
is slightly red-detuned from $U$ and 
blue-detuned. 

Third, if we turn to superconductivity, Floquet physics 
can be used to even convert repulsive interactions into 
attractive ones by applying (linearly-polarised) laser.\cite{Tsuji2011} 
Fourth, we can turn the 
usual pairing into an exotic $\eta$-pairing 
(condensation of pairs at Brillouin 
zone corners) 
by illuminating a superconductor with a 
(linearly-polarised) laser.\cite{Kitamura2016} 
There, the Floquet effective 
pair-hopping is drastically modulated 
when $\omega$ is close to $U$, again 
with different behaviours between $\omega$ 
red-detuned or blue-detuned from $U$.  

Now the purpose of the present paper is to 
encompass topological and superconducting 
properties to seek whether a ``Floquet-induced 
topological superconductivity" can arise.  
While there have been various attempts at realising Floquet topological superconducting states~\cite{Ezawa2014,Zhang2015,Takasan2017,Chono2020,Kumar2021,Dehghani2021}, 
an obstacle is the fact that the pairing symmetry 
is hard to be controlled in a direct manner, since the gap function does not couple to electromagnetic fields.  
Thus, usually, one has to modulate the one-body part in a nontrivial manner (with the gap function kept intact), which e.g. necessitates the Rashba spin-orbit coupling. 
Instead, the strategy of the present paper 
is to exploit the peculiar \textit{laser-induced 
interactions emergent in strongly-correlated systems}.
Namely, we illuminate a circularly-polarised light (CPL) 
to the repulsive Hubbard model in the strong-coupling regime to 
modulate the pairing interaction and the resultant gap function.
We shall show that a $d_{x^2-y^2}$-wave superconductor is indeed changed into a topological $d_{x^2-y^2}+id_{xy}$ wave [Fig.~\ref{fig:overview}f].  
This will be shown in the Floquet formalism for the Gutzwiller-projected 
effective Hamiltonian with the time-periodic Schrieffer-Wolff transformation.  
The $d+id$ pairing is shown to arise from the 
chiral spin coupling along with the three-site term 
caused by the CPL [Fig.~\ref{fig:overview}e].  The latter 
term turns out to remain significant 
even for low frequencies and low intensities of the CPL.  
This will be reflected in a phase diagram against the laser intensity 
and temperature obtained here for various frequencies red-detuned from 
the Hubbard $U$, along with transient dynamics.  
 

\section{Results}

\subsection{Low-energy effective Hamiltonian}

We take in the present study the hole-doped Mott insulator on a square lattice, which is irradiated by an intense circularly-polarised light (CPL).
This can be minimally modelled by the repulsive 
Hubbard model, with a Hamiltonian 
\begin{align}
\hat{H}(t)&=-\sum_{ij\sigma}t_{ij}e^{-i\bm{A}(t)\cdot\bm{R}_{ij}}\hat{c}_{i\sigma}^\dagger\hat{c}_{j\sigma}+\frac{U}{2}\sum_i \hat{n}_{i}(\hat{n}_{i}-1),\label{eq:Hubbard}
\end{align}
where $\hat{c}_{j\sigma}$ annihilates an electron on the site $j$ at $\bm{R}_j$ with spin $\sigma=\uparrow,\downarrow$, and $\hat{n}_i=\sum_{\sigma}\hat{n}_{i\sigma}=\sum_{\sigma}\hat{c}_{i\sigma}^\dagger\hat{c}_{i\sigma}$ is the density operator. 
For the hopping amplitude $t_{ij}$, here 
we consider the second-neighbour ($t_0^\prime$) as well 
as the nearest-neighbour hopping ($t_0$), as 
necessitated for realising a CPL-induced 
topological superconductivity 
[as we shall see below Eq.~(\ref{eq:delta-f})].  
The onsite repulsion is denoted as $U\;(>0)$. 

The circularly-polarised laser field is introduced via the Peierls phase, which involves 
$\bm{R}_{ij}\equiv\bm{R}_{i}-\bm{R}_{j}$, 
and the vector potential, 
\begin{align}
\bm{A}(t)&=\frac{\bm{E}}{2i\omega}e^{-i\omega t}-\frac{\bm{E}^\ast}{2i\omega}e^{i\omega t},\\
\bm{E}&=E(1,i),
\end{align}
for the right-circularly polarised case; replace $\bm{E}$ with $\bm{E}^\ast$ for the left circulation.   We set $\hbar=e=1$ hereafter.

Since the laser electric field 
is time-periodic, $\hat{H}(t+2\pi/\omega)=\hat{H}(t)$,
we can employ the Floquet formalism~\cite{Oka2019}, where the eigenvalue, called quasienergy, of the discrete time translation plays a role of energy. 
Because the quasienergy spectrum is invariant under the time-periodic unitary transformation, $e^{-i\hat{\Lambda}(t)}$ with $\hat{\Lambda}(t+2\pi/\omega)=\hat{\Lambda}(t)$,
we can introduce an effective static Hamiltonian $\hat{H}_{\text{F}}$ as 
\begin{align}
\hat{H}_{\text{F}} = e^{i\hat{\Lambda}(t)}[\hat{H}(t)-i\partial_t]e^{-i\hat{\Lambda}(t)},\label{eq:canonical}
\end{align}
where $\hat{\Lambda}(t)$ is determined such that $\hat{H}_{\text{F}}$ becomes time-independent. 
With such a transformation, we can analyse the time-dependent original problem using various approaches that are designed for static Hamiltonians. 

While it is difficult to determine $\hat{\Lambda}(t)$ and $\hat{H}_{\text{F}}$
in an exact manner, 
we can obtain their perturbative expansions in various situations.  
The high-frequency expansion~\cite{Bukov2015,Eckardt2015,Mikami2016} is a seminal example of such perturbative methods, where the effective Hamiltonian $\hat{H}_{\text{F}}$ is obtained as an asymptotic series in $1/\omega$, and widely used for describing the Floquet topological phase transition.
However, since the onsite interaction $U$ in Mott insulators is typically much greater than the photon energy $\omega$, 
the high-frequency expansion is expected to be invalidated for the present case. 

Still, in the present system in a strongly-correlated regime $t_0\ll U$, we can employ the Floquet extension of the strong-coupling expansion (Schrieffer-Wolff transformation)~\cite{Mentink2014,Bukov2016,Kitamura2017,Claassen2017,Kumar2021}, when the photon energy $\omega$ exceeds the band width $\sim t_0$ of the doped holes (and chosen to be off-resonant with the Mott gap $\sim U$), i.e., $t_0\ll\omega< U$. As we describe the detailed derivation in Methods, we can obtain $\hat{\Lambda}(t)$ and $\hat{H}_{\text{F}}$ in perturbative series in the hopping amplitudes $t_0,t_0^\prime$.
While the $t_0\ll U$ assumption yields a low-energy effective model known in the undriven case as a $t$-$J$ model~\cite{Ogata2008} which describes the dynamics of holes in the background of localised spins, an essential difference in the Floquet states under CPL is that extra terms emerge.  
Namely, the Hamiltonian of the irradiated case reads
\begin{align}
\hat{H}_{\text{F}} & =-\sum_{ij\sigma}\tilde{t}_{ij}\hat{P}_{G}\hat{c}_{i\sigma}^{\dagger}\hat{c}_{j\sigma}\hat{P}_{G}
+\frac{1}{2}\sum_{ij}J_{ij}\left(\hat{\bm{S}}_{i}\cdot\hat{\bm{S}}_{j}-\frac{1}{4}\hat{n}_{i}\hat{n}_{j}\right)
\nonumber \\
 & +\sum_{ijk\sigma\sigma^{\prime}}\Gamma_{i,j;\,k}\hat{P}_{G}\left[(\hat{c}_{i\sigma}^{\dagger}\bm{\sigma}_{\sigma\sigma^{\prime}}\hat{c}_{j\sigma^{\prime}})\cdot\hat{\bm{S}}_{k}-\frac{1}{2}\delta_{\sigma\sigma^{\prime}}\hat{c}_{i\sigma}^{\dagger}\hat{c}_{j\sigma}\hat{n}_{k}\right]\hat{P}_{G}
 \nonumber \\
& +\frac{1}{6}\sum_{ijk}J^\chi_{ijk}(\hat{\bm{S}}_{i}\times\hat{\bm{S}}_{j})\cdot\hat{\bm{S}}_{k}
 ,\label{eq:tjjchi}
\end{align}
where $\hat{P}_G=\prod_i(1-\hat{n}_{i\uparrow}\hat{n}_{i\downarrow})$ projects out doubly-occupied configurations. Here, we have retained all the processes up to the second order (in the hopping amplitudes $t_0,t_0^\prime$), and additionally taken account of the fourth-order processes for the last line of the above equation.  Thus 
the photon-modified exchange interaction, $J_{ij}$, 
and the photon-induced correlated hopping, $\Gamma_{i,j;\,k}$, 
are of second order, while the photon-generated 
chiral spin coupling, 
$J^{\chi}_{ijk}$, is of fourth order.  

Let us look at the effective model in the laser field term by term.  
First, the hopping amplitude of the holes is renormalised from $t_{ij}$ into $\tilde{t}_{ij}=t_{ij}\mathcal{J}_0(A_{ij})$ 
due to the time-averaged Peierls phase~\cite{Dunlap1986,Eckardt2009}, where $A_{ij}=E|\bm{R}_{ij}|/\omega$ and $\mathcal{J}_m$ is the $m$-th Bessel function.  
Then we come to the interactions.  The spin-spin interaction 
has a coupling strength dramatically affected 
by the intense electric field, since it is mediated by kinetic motion of electrons. Namely, the static kinetic-exchange interaction $J$ is modulated as~\cite{Mentink2014}
\begin{align}
J_{ij}= \sum _{m=-\infty}^\infty \frac{4t_{ij}^2\mathcal{J}_m(A_{ij})^2}{U-m \omega},\label{eq:heisenberg}
\end{align}
which is a sum over $m$-photon processes, and 
appears in the second term on the first line of Eq.~(\ref{eq:tjjchi}).

\begin{figure}[t]
\begin{centering}
\includegraphics[width=0.9\linewidth]{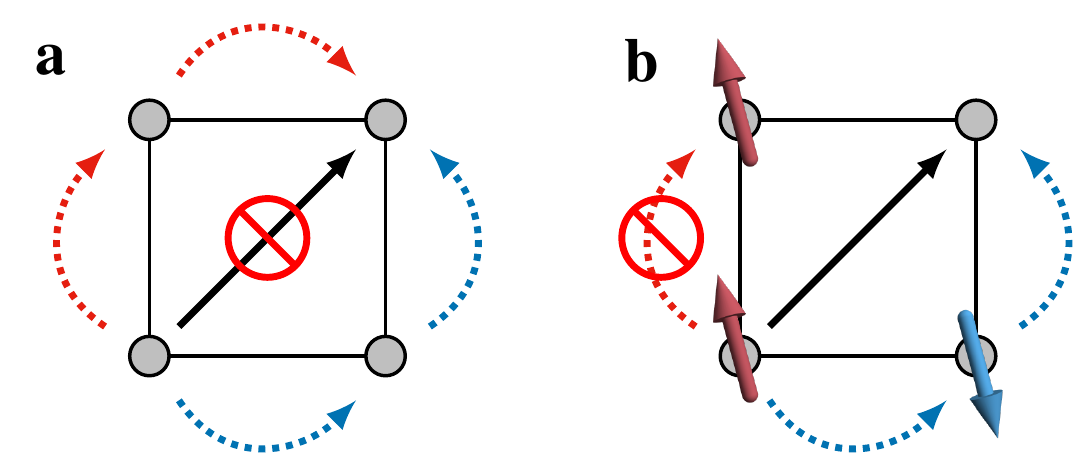}
\par\end{centering}
\caption{\label{fig:two-step}\textbf{Schematic picture for two-step hopping.} 
\textbf{a} Photon-induced hopping processes in the noninteracting square lattice, where contributions from two different paths (red and blue dotted arrows) cancel with each other. \textbf{b} For a strongly-correlated model, 
the photon-induced hopping processes are sensitive to the charge and spin configurations on the path, which lifts the cancellation in general.}
\end{figure}

While these modulations of $t$ and $J$ also occur for the case of linearly-polarised lasers (i.e., for time-reversal symmetric modulations), 
a notable feature of the CPL is the emergence of the \textit{time-reversal breaking many-body interactions} 
specific to the strong correlation. 
In the one-body Floquet physics, the two-step hopping (i.e., a perturbative process with hopping twice) in CPL 
becomes imaginary (and thus breaks the time-reversal symmetry) 
already on the noninteracting level if we take a honeycomb lattice~\cite{Oka2009,Kitagawa2011,Mikami2016}, 
while such an imaginary hopping does not arise for the (noninteracting) square lattice due to a cancellation of contributions from different paths as depicted in Fig.~\ref{fig:two-step}a, which is why a honeycomb 
lattice fits with the FTI.   

If we move on to correlated systems, 
on the other hand, the hopping process involves interactions for spinful 
electrons on the path, 
which lifts such cancellations, as exemplified in Fig.~\ref{fig:two-step}b.  
This implies that such a time-reversal breaking term is present 
in (interacting) square lattice as well.  
The two-step \textit{correlated} hopping $\Gamma$ appears in the second-order perturbation in the presence of holes [the second line of Eq.~(\ref{eq:tjjchi})], which 
involves three sites and sometimes referred to as the ``three-site term"~\cite{Ogata2008}. 
While the three-site term also appears in the undriven case with the amplitude $\Gamma_{i,j;\,k}=J_{ij}/4$, 
this term involving the dynamics of holes (i.e., the change of holes' positions) 
is usually neglected because its contribution is small in the low-doping regime [see the Gutzwiller factor in the next subsection, $\sim g^2\delta$ in Eqs.~(\ref{eq:dispersion}), (\ref{eq:gapfunction})].  
In sharp contrast, when the system is driven by a CPL, $\Gamma_{i,j;\,k}$ acquires an important 
\textit{three-site imaginary part} as
\begin{align}
\text{Im}\,\Gamma_{i,j;\,k} & =\sum_{m=1}^\infty\frac{2t_{ik}t_{kj}\mathcal{J}_{m}(A_{ik})\mathcal{J}_{m}(A_{kj})}{m\omega(1-m^2\omega^2/U^2)}\sin m(\theta_{jk}-\theta_{ik})\label{eq:imgamma}
\end{align}
with $\theta_{ij}$ defined as $\bm{R}_{ij}=|\bm{R}_{ij}|(\cos\theta_{ij},\sin\theta_{ij})$. 
The term triggers a topological superconductivity even 
when it is small, as we shall see.

In addition to the two-step correlated hopping $\Gamma$, another 
time-reversal breaking term arises 
if we go over to higher-order perturbations.
In the previous studies of the half-filled case~\cite{Kitamura2017,Claassen2017}, 
an emergent three-spin interaction (i.e., the scalar \textit{spin chirality term} $J^\chi$) is shown 
to appear in the fourth-order perturbation. 
This appears on the last line 
in Eq.~(\ref{eq:tjjchi}) with a coefficient $J^\chi_{ijk}$.  
While this term is basically much smaller than the second-order terms, 
it may become comparable with the contribution from $\text{Im}\,\Gamma$, as the spin chirality term does not accompany the dynamics of holes
and has a larger Gutzwiller factor [$\sim g^3$ in Eqs.~(\ref{eq:dispersion}), (\ref{eq:gapfunction})].

\subsection{Mean-field decomposition under Gutzwiller ansatz}

\begin{figure*}[t]
\begin{centering}
\includegraphics[width=0.8\linewidth]{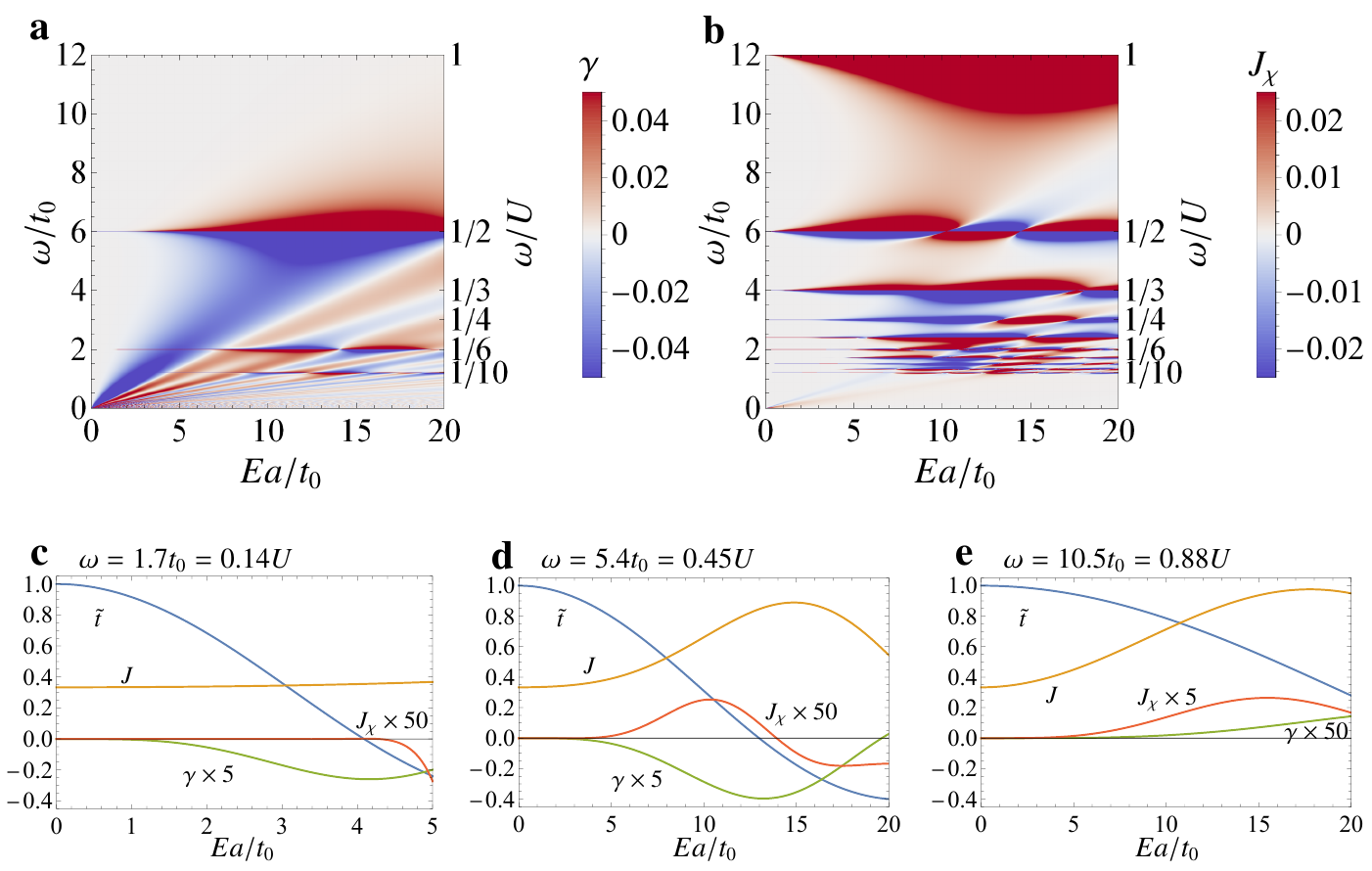}
\par\end{centering}
\caption{\label{fig:params}\textbf{Dependence of the coupling constants on the laser intensity for the driven model.} 
\textbf{a}, \textbf{b} Two-step correlated hopping $\gamma$ (\textbf{a}), and scalar spin-chirality term $J_\chi$ (\textbf{b}) against circularly-polarised-light amplitude $E$ and driving frequency $\omega$. 
Fig.~\ref{fig:overview}d corresponds to a cross section of \textbf{b} at $E=6t_0/a$.  
\textbf{c}-\textbf{e} The renormalised hopping amplitude $\tilde{t}$, exchange coupling $J$, along with 
the two-step correlated hopping $\gamma$ and the chiral spin-coupling $J_\chi$, against $E$. The driving frequency is chosen as 
$\omega=1.7t_0=0.14U$ (\textbf{c}), $\omega=5.4t_0=0.45U$ (\textbf{d}), or $\omega=10.5t_0=0.88U$ (\textbf{e}) with 
$t_0$: bare hopping amplitude, $a$: lattice constant. 
Note a difference in the horizontal scale between \textbf{c} and \textbf{d,e}.  
We take here the bare second-neighbour hopping $t_0^\prime=-0.2t_0$, the onsite interaction $U=12t_0$, 
and a doping level $\delta=0.2$.
Here, the infinite summations of Fourier components [$\sum_m$ in Eq.~(\ref{eq:imgamma}) and $\sum_{lmn}$ in Eq.~(\ref{eq:chirality})] are truncated such that the Taylor series up to $E^{20}$ is reproduced. This is accurate enough for $\omega> t_0$, while for the low-frequency regime $\omega< t_0$ an intricate resonant structure due to the 
$\omega=U/$integer resonance (seen also for $t_0<\omega<4t_0$) is not captured.
}
\end{figure*}

Let us investigate the fate of the $d$-wave superconductivity in the presence of the time-reversal breaking terms.
To this end, we here adopt the Gutzwiller ansatz~\cite{Ogata2003,Ogata2008} for the ground-state wavefunction and perform a mean-field analysis, as elaborated in Methods.
After the approximate evaluation of the Gutzwiller projection, the problem of determining the ground state is turned into an energy minimisation of Eq.~(\ref{eq:tjjchi}) without the Gutzwiller factor $\hat{P}_G$ if we renormalise the parameters.
The mean-field solution is then formulated as a diagonalisation of the the Bogoliubov-de Gennes Hamiltonian in the momentum space,
\begin{align}
\hat{H}_{\text{F}} & =\sum_{\bm{k}}\begin{pmatrix}\hat{c}_{\bm{k}\uparrow}\\
\hat{c}_{-\bm{k}\downarrow}^{\dagger}
\end{pmatrix}^{\dagger}\begin{pmatrix}\varepsilon(\bm{k}) & F(\bm{k})\\
F(\bm{k})^{\ast} & -\varepsilon(-\bm{k})
\end{pmatrix}\begin{pmatrix}\hat{c}_{\bm{k}\uparrow}\\
\hat{c}_{-\bm{k}\downarrow}^{\dagger}
\end{pmatrix}\label{eq:BdG}\\
 & =\sum_{\bm{k}}\begin{pmatrix}\hat{c}_{\bm{k}\uparrow}\\
\hat{c}_{-\bm{k}\downarrow}^{\dagger}
\end{pmatrix}^{\dagger}\left[\sum_{\tau}\begin{pmatrix}\varepsilon_\tau & F_\tau\\
F_\tau^{\ast} & -\varepsilon_\tau
\end{pmatrix}\cos\bm{k}\cdot\bm{R}_{i,i+\tau}\right]
\begin{pmatrix}\hat{c}_{\bm{k}\uparrow}\\
\hat{c}_{-\bm{k}\downarrow}^{\dagger}
\end{pmatrix}.
\end{align}
Here we have
\begin{align}
\varepsilon_\tau & =-g\delta\tilde{t}_{i,i+\tau}-g^2\frac{3-\delta^{2}}{8}J_{i,i+\tau}\,\chi_{\tau}
\nonumber\\
& -g^2\delta\sum_{\tau^{\prime}}\text{Re}\left(\frac{1-\delta^{2}}{4}\Gamma_{i,i+\tau;\,i+\tau^{\prime}}+\frac{3-\delta}{2}\Gamma_{i,i+\tau+\tau^{\prime};\,i+\tau}\,\chi_{\tau^{\prime}}\right) \nonumber \\
&-\frac{3}{16}g^3\sum_{\tau^{\prime}}J^{\chi}_{i,i+\tau,i+\tau+\tau^{\prime}}\text{Im}(\Delta_{\tau+\tau^{\prime}}^{\ast}\Delta_{\tau^{\prime}}),\label{eq:dispersion}
\\ 
F_\tau  &=g^{2}\frac{3+\delta^{2}}{8}J_{i,i+\tau}\Delta_{\tau}+g^{2}\delta\frac{3+\delta}{2}\sum_{\tau^{\prime}}\text{Re}\Gamma_{i,i+\tau+\tau^{\prime};\,i+\tau}\Delta_{\tau^{\prime}}\nonumber\\
 & +ig^{2}\sum_{\tau^{\prime}}\left(\delta\frac{3+\delta}{2}\text{Im}\Gamma_{i,i+\tau+\tau^{\prime};\,i+\tau}+\frac{3}{8}gJ_{\chi,i,i+\tau,i+\tau+\tau^{\prime}}\,\chi_{\tau+\tau^{\prime}}\right)\Delta_{\tau^{\prime}},
\label{eq:gapfunction}
\end{align}
where the expressions are independent of the site index $i$ due to the spatial translational symmetry, and we have denoted the doping level as 
\begin{equation}
\delta=1-\frac{1}{N}\sum_i\langle\hat{n}_i\rangle
\end{equation}
with $N$ being the number of lattice sites.  
Here $\tau$ runs over $\tau=mx+ny$ with $m,n\in\mathbb{Z}$, and the label $i+\tau$ represents the site at $\bm{R}_{i+\tau}=\bm{R}_i+(m,n)$ in units of the lattice constant.  
The dependence on $\delta$ and a factor $g=2/(1+\delta)$ appears as a result of the Gutzwiller projection, which suppresses the contribution from charge dynamics as represented by $\tilde{t}$ and $\Gamma$ in the small $\delta$ regime. 
For the detailed derivation, see Methods.  
Throughout the present study, we take $\delta=0.2$.  
The mean-field Hamiltonian is self-consistently determined 
for the bond order parameter $\chi$ and 
the pairing amplitude $\Delta$ as
\begin{align}
\chi_{\tau}&=\frac{1}{N}\sum_i\langle\hat{c}_{i\uparrow}^{\dagger}\hat{c}_{i+\tau\uparrow}+\hat{c}_{i\downarrow}^{\dagger}\hat{c}_{i+\tau\downarrow}\rangle,\\
\Delta_{\tau}&=\frac{1}{N}\sum_i\langle\hat{c}_{i\uparrow}\hat{c}_{i+\tau\downarrow}-\hat{c}_{i\downarrow}\hat{c}_{i+\tau\uparrow}\rangle.
\end{align}

\begin{figure*}[t]
\begin{centering}
\includegraphics[width=\linewidth]{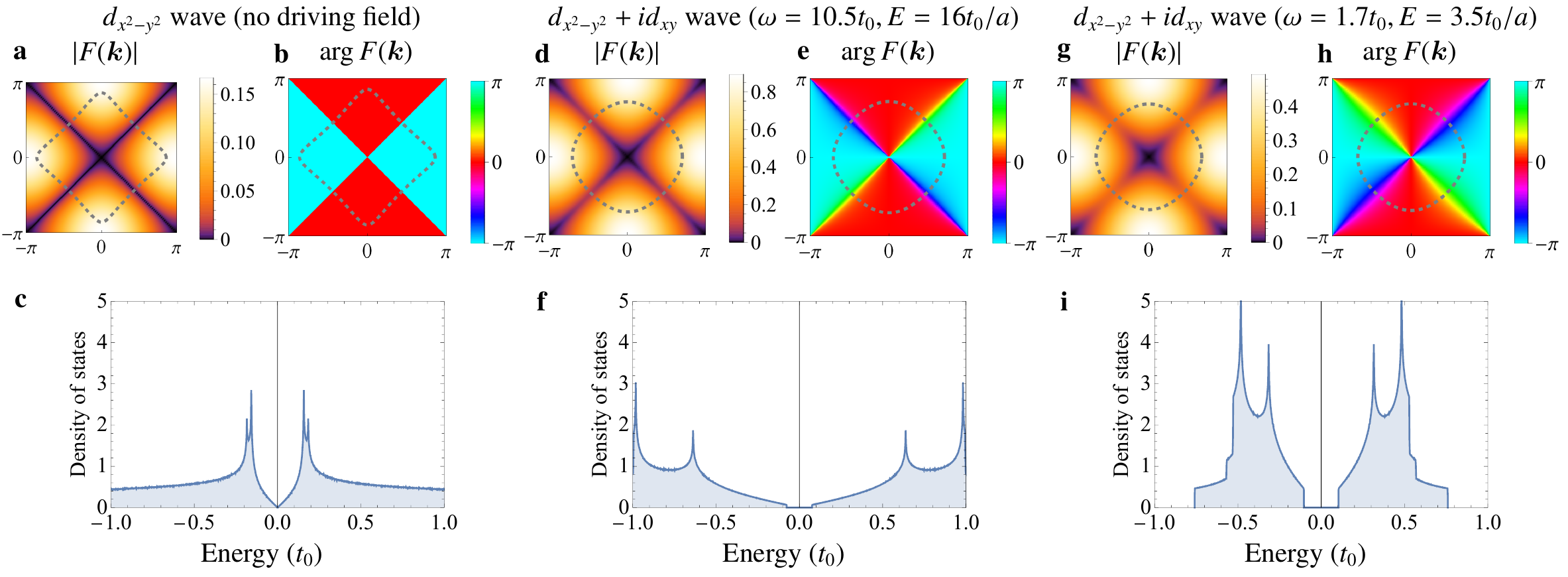}
\par\end{centering}
\caption{\textbf{Gap functions.} \textbf{a},\textbf{d},\textbf{g} Absolute values of the gap function, $|F(\bm{k})|$. \textbf{b},\textbf{e},\textbf{h} Phases, $\arg F(\bm{k})$. 
 The dashed lines represent the Fermi surface at the critical temperature $T=T_c$. 
\textbf{c},\textbf{f},\textbf{i} Density of states against 
energy in units of the bare hopping amplitude $t_0$. These are shown for \textbf{a}-\textbf{c} $d_{x^2-y^2}$-wave superconductivity in the absence of driving field, and \textbf{d}-\textbf{i} $d_{x^2-y^2}+id_{xy}$-wave superconductivity in the circularly-polarised laser.  The driving frequency $\omega$ and the field amplitude $E$ are chosen as $\omega=10.5t_0=0.88U, E=16t_0/a$ (\textbf{d}-\textbf{f}), or $\omega=1.7t_0=0.14U, E=3.5t_0/a$ (\textbf{g}-\textbf{i}).  
We take a parameter set: the bare second-neighbour hopping $t_0^\prime=-0.2t_0$, the onsite interaction $U=12t_0$, 
and a doping level $\delta=0.2$.}
\label{fig:gap}
\end{figure*}

In the absence of the external field, the system undergoes a phase transition from normal to the $d_{x^2-y^2}$-wave superconductivity when the temperature is sufficiently low~\cite{Ogata2008}. The corresponding order parameter is 
\begin{equation}
\Delta_{\pm x}=-\Delta_{\pm y}\equiv \Delta,
\end{equation}
for which the gap function $F(\bm{k})$ becomes 
\begin{equation}
F^{x^2-y^2}(\bm{k})=\frac{3}{4}g^2J\Delta(\cos k_x-\cos k_y),
\end{equation}
as derived from the first term in Eq.~(\ref{eq:gapfunction}) with $\tau=\pm x,\pm y$. Here we have dropped a small correction due to $\delta$. 
Without a loss of generality, we assume that $\Delta$ is real.
The $d_{x^2-y^2}$ gap function has nodal lines along $k_y=\pm k_x$, across which the gap function changes sign (See Figs.~\ref{fig:gap}a-\ref{fig:gap}c below).

Now let us look into the possibility for the laser field 
converting this ground state into \textit{topological superconductivity} from the time-reversal breaking terms [the second line in Eq.~(\ref{eq:gapfunction})]. 
The leading term should be those for $\tau^\prime=\pm x,\pm y$ (with $\Delta_{\tau^\prime}=\pm\Delta$), which is nonzero even for the original $d_{x^2-y^2}$-wave ansatz and results in an imaginary gap function $\propto i\Delta$. 
In particular, gathering the terms with $\tau=\pm(x+y),\pm(x-y)$, we obtain the leading modulation to the gap function as 
\begin{align}
F^{xy}(\bm{k}) \simeq 3i
g^{2}\left[4\delta\gamma+g(J_{\chi}\,\chi_{x}+J_{\chi}^{\prime}\,\chi_{2x+y})\right]\Delta\sin k_{x}\sin k_{y}\label{eq:delta-f}
\end{align}
where we have defined 
\begin{gather}
\gamma = \text{Im}\,(\Gamma_{i-x,i;\,i+y}
-\Gamma_{i-x-y,i+x;\,i}),\\
J_{\chi} = J_{i,i+y,i+x}^{\chi}\;, \; \;
J_{\chi}^{\prime} = J_{i-x,i,i+x+y}^{\chi}.
\end{gather}
Thus we are indeed led to 
an emergence of a topological (chiral) $d_{x^2-y^2}+id_{xy}$ superconductivity with 
$F(\bm{k})\simeq F^{x^2-y^2} +iF^{xy}$
with a full gap.  For the full expression of the modulated gap function, see Eq.~(\ref{eq:gap-full}) in Methods.
The key interactions, $\gamma$ and $J_\chi$, take nonzero values 
when the original Hubbard Hamiltonian Eq.~(\ref{eq:Hubbard}) has the second-neighbour hopping $t_0^\prime$:
Since the two-step correlated hopping $\Gamma_{i,j;\,k}$ is composed of two hoppings,  $k\to i$ and $j\to k$,
we can see that the imaginary coefficient above has 
$\gamma\propto t_0t_0^\prime$.
The chiral spin-coupling $J_\chi\propto (t_0t_0^\prime)^2$ also necessitates the next-nearest-neighbour hopping, 
which can be deduced from Eq.~(\ref{eq:chirality-details}) in Methods.


Having revealed the essential couplings for the chiral superconductivity, 
let us now explore how we can 
\textit{optimise} the field amplitude and frequency for realising larger topological gaps.  For this, 
we set here $t_0^\prime=-0.2t_0$ and $U=12t_0$ 
having cuprates with $t_0\simeq0.4$~eV in mind as a typical example.

We plot the essential $\gamma$ and $J_\chi$ against the driving amplitude $E$ and frequency $\omega$ in Figs.~\ref{fig:params}a and \ref{fig:params}b. 
If we look at the overall picture, 
we can see that the dynamical time-reversal breaking is strongly enhanced along some characteristic frequencies in 
a \textit{resonant} fashion at $\omega/U = 1/$integer, which we can capture from the expressions for the coupling constants in the small-amplitude 
regime as follows.

The two-step correlated hopping $\gamma$ is given, 
in the leading order in the amplitude, as 
\begin{equation}
\gamma=\frac{2t_{0}t_{0}^{\prime}\mathcal{J}_{2}\left(\frac{Ea}{\omega}\right)\mathcal{J}_{2}\left(\sqrt{2}\frac{Ea}{\omega}\right)}{\omega(1-4\omega^{2}/U^{2})}+O(E^{12}).
\end{equation}
This expression, as a function of $E$, 
takes the maximal value around $E \simeq 2.45\omega/a$, 
while diverges for $\omega\to0$ or $\omega\to U/2$ for each 
value of $E$,
as seen from the energy denominator.  
The enhancement occurs even 
in the low-frequency regime, which should be advantageous 
for experimental feasibility, 
since the required field amplitude ($E\simeq 2.4\omega/a$) 
can be small. 

If we turn to the spin-chirality term, $J_\chi$ has a complicated form involving Bessel functions [see Eq.~(\ref{eq:chirality-details}) below]. If we 
Taylor-expand it in $E$, we have
\begin{equation}
J_{\chi}\sim\frac{2(Ea)^4 t_{0}^{2}t_{0}^{\prime2}(2U^6+75\omega^{2}U^4-399\omega^{4}U^2-164\omega^{6})}{\omega(U^2-\omega^{2})^{3}(U^2-4\omega^{2})^{3}},
\label{eq:Jchi-fourth}
\end{equation}
which takes large values around $\omega=U$ due to the 
energy denominator in the above expression, while $\gamma$ is small in this regime.
The expression reveals that a dynamical time-reversal breaking also occurs 
as a fourth-order nonlinear effect with respect to the field strength $E$.

\begin{figure*}[t]
\begin{centering}
\includegraphics[width=\linewidth]{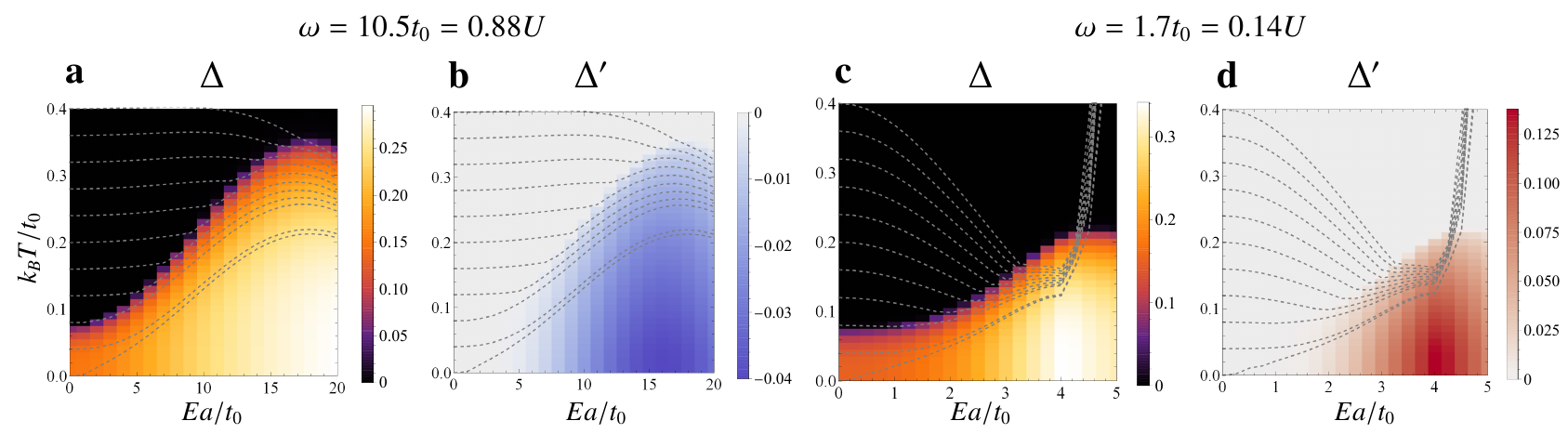}
\par\end{centering}
\caption{\label{fig:phase}\textbf{Finite-temperature phase diagram.} 
The superconducting order parameter $|\Delta|$ (\textbf{a}, \textbf{c}), and the time-reversal breaking order parameter $\Delta^\prime \equiv -i\Delta_{\pm (x+y)}=i\Delta_{\pm (x-y)}$
 (\textbf{b}, \textbf{d}) plotted against the field amplitude $E$ and temperature $T$.
The driving frequency is set to 
 $\omega=10.5t_0=0.88U$ (\textbf{a}, \textbf{b}) or $\omega=1.7t_0=0.14U$ (\textbf{c}, \textbf{d}). Dotted lines represent the effective temperature of the Floquet Hamiltonian when the system is quenched from $E=0$ at a 
temperature that is the leftmost starting point 
of each dotted line. We take the same 
bare second-neighbour hopping $t_0^\prime=-0.2t_0$, onsite interaction $U=12t_0$, 
and doping level $\delta=0.2$ as in Fig.4.}
\end{figure*}

We can see in Figs.~\ref{fig:params}c-e 
how $J_{\chi}$, as well as the exchange interaction 
$J$ and 
the renormalised nearest-neighbour hopping $\tilde{t}$, 
vary with the CPL 
intensity for several representative frequencies
for which the dynamical time-reversal breaking becomes prominent.  
The hopping amplitude $\tilde{t}$ 
vanishes around the peak of $\gamma$ ($E \simeq 2.45\omega/a$), as seen in Figs.~\ref{fig:params}c,d. 
This involves a zero of the Bessel 
function (at $E\sim2.40\omega/a$), and known as the dynamical localisation~\cite{Dunlap1986}.  
In Figs.~\ref{fig:params}d,e, we can see a strong enhancement of $J$.  Panel~d has $\omega = 0.45U$ (slightly red-detuned from $U/2$), while panel~e 
has $\omega = 0.88U$ (slightly red-detuned from $U$).  
If we go back to 
Eq.~(\ref{eq:heisenberg}), the former has to do with 
the energy denominator for $m=2$, while the 
latter for $m=1$. While the driving frequency is set 
to be red-detuned from the resonance (which we can call 
``$U-m\omega$ resonance") in these results, the blue-detuned cases would give negative contributions from these terms, 
which will lead to a ferromagnetic exchange interaction, unfavouring the spin-singlet $d$-wave.  
With an optimal choice of the driving field, $\gamma$ 
attains a significantly large value $\simeq0.04 t_0$, 
which yields $F^{xy}(\bm{k})\simeq 0.3t_0\times i\Delta\sin k_{x}\sin k_{y}$ (for $\delta=0.2$).
This is remarkably large and \textit{comparable} with the 
undriven $F^{x^2-y^2}(\bm{k})\simeq 0.7t_0\times \Delta (\cos k_{x}-\cos k_{y})$, 
even though the coupling constant itself 
is much smaller than $J$.  This is the first key result 
of the present work.


\subsection{Phase diagram}

Now we investigate the ground state of the effective static Bogoliubov-de Gennes Hamiltonian for several choices of the CPL parameters.
Let us show the gap function and the density of states in Fig.~\ref{fig:gap}.  
In the absence of the external field in Figs.~\ref{fig:gap}a-\ref{fig:gap}c, 
the gap function has the $d_{x^2-y^2}$ symmetry with nodal lines that give the zero gap at the Fermi energy.  
We now switch on the CPL, with $\omega=10.5t_0=0.88U$, $E=16t_0/a$ for Figs.~\ref{fig:gap}d-\ref{fig:gap}f, or with $\omega=1.7t_0=0.14U$, $E=3.5t_0/a$ for Figs.~\ref{fig:gap}g-\ref{fig:gap}i.
The former represents the case where 
$J_\chi$ is dominant (see Fig.~\ref{fig:params}e), while in the latter $\gamma$ plays the central role.  
In both cases we have the $d_{x^2-y^2}+id_{xy}$ pairing, where 
the nodal lines are gapped out due to the complex gap function.  
We can indeed see in Figs.~\ref{fig:gap}f,i clear energy gaps in the density of states with a gap size comparable with the original $d_{x^2-y^2}$ superconducting gap (as measured by 
the energy spacing between the two peaks in the 
density of states).  Note that 
the band width and the $d_{x^2-y^2}$ superconducting gap 
are renormalised there, due to the modified $\tilde{t}$ and $J$.

To examine the robustness of the $d+id$ superconductivity, we calculate $|\Delta|$ and $\Delta^\prime \equiv -i\Delta_{\pm (x+y)}=i\Delta_{\pm (x-y)}$ at finite temperatures, 
which gives a phase diagram against $T$ and field amplitude $E$ in Fig.~\ref{fig:phase}. 
We can see that the critical temperature $T_c$, as delineated by the region for $\Delta\neq0$, increases significantly 
as the laser intensity $E$ is increased. The enhanced $T_c$ originates from the fact that $J$ [and $\text{Re}\,\Gamma$; see 
Eq.~(\ref{eq:Gamma}) below] are enhanced when the laser is applied, as we have seen in Figs.~\ref{fig:params}c-\ref{fig:params}e. The time-reversal breaking order parameter $\Delta^\prime$ emerges below $T_c$ down to $T=0$,
with the field dependence emerging from those of $\gamma$ and $J_\chi$.

\begin{figure*}[t]
\begin{centering}
\includegraphics[width=\linewidth]{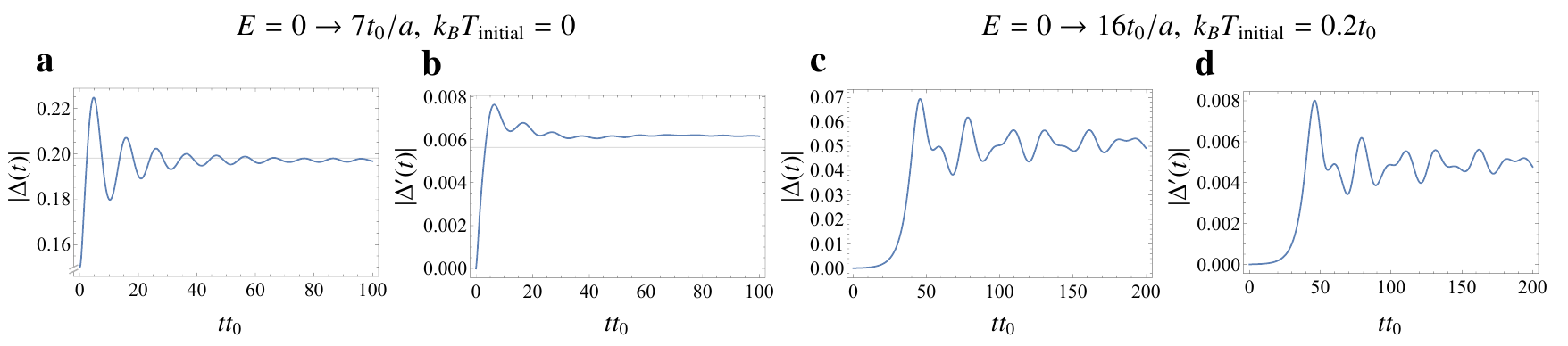}
\par\end{centering}
\caption{\label{fig:dynamics}\textbf{Quench dynamics of the order parameters.} Time evolution of 
$|\Delta|=|\langle c_{i\uparrow} c_{i+x,\downarrow} \rangle|$ (\textbf{a}, \textbf{c}) and $|\Delta^\prime|=|\langle c_{i\uparrow} c_{i+x+y,\downarrow} \rangle|$ (\textbf{b}, \textbf{d})
after a sudden quench of the field amplitude from $E=0$ to $E=7t_0/a$ with an initial temperature $k_BT_{\text{initial}}=0$ (\textbf{a}, \textbf{b}), or to $E=16t_0/a$ with an initial 
$k_BT_{\text{initial}}=0.2t_0$ (\textbf{c}, \textbf{d}).
The latter corresponds to the case in which 
the superconductivity appears even when 
we start from an initial temperature above $T_c$.  
Horizontal lines indicate the equilibrium value at the effective temperature.
The driving frequency is set to $\omega=10.5t_0=0.88U$ with $t_0$ being the bare hopping amplitude.
}
\end{figure*}

\subsection{Transient dynamics}
So far, we have investigated static properties of the effective Hamiltonian $\hat{H}_{\text{F}}$. While the results clearly show the presence of Floquet topological superconductivity in a wide parameter region,
an important question from a dynamical viewpoint is whether we can achieve the Floquet topological superconductivity within 
a short enough time scale, over which the description with the effective Hamiltonian Eq.~(\ref{eq:tjjchi}) is valid (i.e., over which we can neglect 
the coupling to phonons, the heating process due to higher-order perturbations, etc).

Thus let us study the dynamics of the superconducting gap, by solving a quench problem formulated as follows. 
We first prepare the initial state as the mean-field ground state of the equilibrium $t$-$J$ model [i.e., Eq.~(\ref{eq:tjjchi}) with $E=0$], 
and then look at the transient dynamics when the 
CPL electric field is 
suddenly switched on, by changing 
the Hamiltonian at $t=0$ to the effective Hamiltonian $\hat{H}_{\text{F}}$ (\ref{eq:tjjchi}) with $E\neq0$.  
In such a treatment, $\langle\hat{H}_{\text{F}}\rangle$ is a conserved quantity for $t>0$, and 
the driven state is expected to thermalise to the equilibrium state of $\hat{H}_{\text{F}}$ having a temperature that 
corresponds to the internal energy $\langle\hat{H}_{\text{F}}\rangle$ (which we call an effective temperature). 
Note that here we implicitly neglect the fact that $\hat{H}_{\text{F}}$s for $E=0$ and $E\neq0$ are on different frames [specified by $\hat{\Lambda}(t)$].

Before exploring the dynamics, it is important to check 
whether the expected steady state of the quench problem remains superconducting.  We can do this by looking at 
the above-defined effective temperature on the phase diagram as indicated by dotted lines in Fig.~\ref{fig:phase}.  
While the effective temperature rises as we increase the field strength $E$ when we start from temperatures below $T_c$, we can see that 
the effective temperature does \textit{not} exceed $T_c$ in a wide parameter region, which implies that the Floquet topological superconductivity should indeed appear as a steady state of the quench dynamics.
A sudden increase of temperature around $E\sim 4t_0/a$ in Figs.~\ref{fig:phase}c, \ref{fig:phase}d is 
due to the band flipping (sign change of $\tilde{t}$), which makes the kinetic energy of the initial state quite large. As we further increase the field strength, the effective temperature eventually reaches infinity and will even become negative~\cite{Tsuji2011}. 

Now, the question is whether the equilibration (known as the Floquet prethermalisation) occurs 
fast enough (e.g., faster than the neglected heating processes, 
within the experimentally accessible pulse duration, etc.). 
We compute the time evolution of the superconducting order parameter in the time-dependent mean-field approximation (with the Gutzwiller ansatz),
and plot the time evolution of $|\Delta|$ and $|\Delta^\prime|$ in Figs.~\ref{fig:dynamics}a, \ref{fig:dynamics}b, where we set the initial temperature at $k_BT_{\text{initial}}=0$ with  $E=7t_0/a$, $\omega=10.5t_0=0.88U$.
Here, the unit of time $\hbar/t_0$ corresponds to $\sim1$~fs for $t_0\simeq0.4$~eV.  
We can see that $|\Delta|$ rapidly evolves with an overshooting behaviour, and converges to a certain value with a damped oscillation.  
Similar behaviour can be found in a previous study of the quench problem for the $d$-wave (but within the $d_{x^2-y^2}$ pairing) superconductor~\cite{Peronaci2015}.  
The time-reversal breaking order parameter $|\Delta^\prime|$ also quickly converges to a nonzero value. This is the second key result in the present work.
We note that the obtained steady state slightly deviates from the equilibrium state with respect to $\hat{H}_{\text{F}}$, where the deviation arises because the pair-breaking scattering~\cite{Peronaci2015} is lacking in the mean-field treatment of the dynamics. 

The curious oscillation in the amplitude of the gap function can be interpreted as an excitation of the \textit{Higgs modes} in superconductors~\cite{Matsunaga2014,Shimano2020,Katsumi2018}, 
and thus the typical time scale for the oscillation (and the emergence of $\Delta^\prime$) can be roughly estimated
as the inverse of the superconducting gap ($\sim1/2|F(\bm{{k}})|$ with an appropriate $k$-average).  
This should be much faster than the time scale for heating, although its nonempirical evaluation would be difficult.
We can analyse the present quench dynamics in a linearised form~\cite{Schwarz2020} if the change in the coupling constant is small,
which reveals that the appearance of $\Delta^\prime$ is described as the $A_{2g}$ Higgs mode in the $d$-wave sector with an amplitude proportional to $E^4$.

If we have a closer look at the effective temperature in Fig.~\ref{fig:phase}, we find an intriguing 
phenomenon: the superconducting state can appear even when 
we start from an initial temperature that is \textit{above} the $T_c$ before laser illumination. 
Namely, some dotted lines that start from $T$ above $T_c$ 
at $E=0$ do plunge into the superconducting 
region as $E$ is increased. We show
the time evolution of the order parameter for this 
``nonequilibrium-induced superconductivity" in Figs.~\ref{fig:dynamics}c, \ref{fig:dynamics}d, where we set $k_BT_{\text{initial}}=0.2t_0$ as an initial temperature, and choose $E=16t_0/a$, $\omega=10.5t_0$. 
Since the homogeneous mean-field ansatz without fluctuations cannot describe
the spontaneous symmetry breaking, 
we instead inspect the growth of a tiny (homogeneous) perturbation $\Delta=10^{-4}$ on the initial state.  
After the quench, both of the gap functions 
$\Delta$ and $\Delta'$ grow exponentially and converge 
respectively to nonzero values, 
although the damped oscillation is slower than the previous case, and the converged values are far below those ($|\Delta|\simeq0.18$ and $|\Delta^\prime|\simeq0.03$) expected for equilibrium with the effective temperature. 


\section{Discussion}
In the present paper we reveal that Floquet-induced 
interactions do indeed give rise to a 
novel way for creating topological superconductivity. 
Let us recapitulate advantages of the present proposal for the Floquet topological superconductivity.
In the previous study of Floquet topological superconductivity in cuprates~\cite{Takasan2017}, 
the topological transition is triggered by the modulation of the kinetic part of the Hamiltonian [diagonal components in Eq.~(\ref{eq:BdG})], with the pairing symmetry of the gap function remaining the same.
The nontrivial structure of the kinetic part necessitates the presence of a strong Rashba spin-orbit coupling, 
which would limit the applicable class of materials.
In the present approach, by contrast, we exploit the correlation effects themselves, 
with which we can directly modulate the pairing symmetry to obtain the topological superconductivity.
Our approach does not require tailored structures in the one-body part, either, 
and thus a simple square-lattice Hubbard model suffices for inducing the topological transition.
The present approach is also advantageous in achieving a large topological gap.  Namely, 
the topological gap we conceive is comparable with $k_BT_c$ (since the $F^{x^2-y^2}(\bm{k})$ and $F^{xy}(\bm{k})$ components have the same order of magnitude), while in the previous studies the size of the topological gap is bounded e.g. by the Rashba spin-orbit coupling and estimated to be $\sim1$~K.

While we have evaluated the increase of the effective temperature upon a sudden change in the field amplitude (which might be evaded by adiabatic ramping of the field), 
there is also a many-body heating process that is dropped in the present formulation.
In isolated, nonintegrable many-body systems in general, the exact eigenstates of $\hat{H}_{\text{F}}$ in the thermodynamic limit  at long times are believed to represent a featureless, infinite-temperature state 
(sometimes called the Floquet eigenstate thermalisation hypothesis~\cite{DAlessio2014,Lazarides2014,Lazarides2014-2,Bukov2016-2,Seetharam2018,Mori2018}).  
Thus, the present strong-coupling expansion for the nontrivial structure should be interpreted as an asymptotic expansion (with a vanishing radius of convergence) of this featureless Hamiltonian.  
The expanded Hamiltonian truncated at an optimal order 
accurately describes dynamics towards a long-lived state (dubbed as Floquet prethermalisation~\cite{Abanin2017,Mori2016,Kuwahara2016,Mori2018}), while the inevitable small truncation error separately describes a slow heating process towards the infinite temperature.

Specifically, the present expansion is characterised by the energy denominators of the form $(nU-m\omega)$ with integers $n,m$, in which 
the enhancement of $\gamma$ and $J_\chi$ grows as 
the resonance is approached.  We have to note, however, that the 
accuracy of the asymptotic expansion becomes degraded, 
i.e., the time scale over which the effective Hamiltonian 
remains valid shrinks (with heating becoming faster) as we come closer to the resonance with vanishing denominators. 
Thus, we will have to 
examine whether the time scale $t\sim10/t_0$ for the emergence of $\Delta^\prime$ (see Fig.~\ref{fig:dynamics}b) 
lies 
within the validity of the Floquet $t$-$J$ description, 
with more sophisticated methods including direct analyses of a time-dependent problem, in future works. 
While the emergence of the topological superconductivity is verified here with a numerically economical approach, 
we do obtain in the present paper 
the effective Hamiltonian with the crucial 
time-reversal breaking by 
fully taking account of the noncommutative nature of the Gutzwiller projector $\hat{P}_\text{G}$ 
(whereas the usual hopping terms are commutative). 
It should be thus important to retain such correlation effects in performing sophisticated calculations. 

As for ``nonequilibrium-induced superconductivity" 
(i.e., laser illumination making 
a system superconducting even when we start from $T$ above $T_c$) discussed in the present paper, 
there is existing literature that reports laser-induced 
phenomena~\cite{Budden2021} along with related theories~\cite{Knap2016,Babadi2017,Murakami2017,Kennes2019}, 
although the time scale and proposed mechanism are 
quite different from the present paper.
Whether the present theory has some
possible relevance will be a future problem.

Let us turn to the required field amplitude and frequency for experimental feasibility.  
Specifically, if we want to employ the enhancement of time-reversal breaking around $\omega=U/2$ or $\omega=U$, 
the required field intensity is $E\sim 10t_0/a$, which 
corresponds to $E\sim10^2$~MV/cm for the typical 
cuprates with $t_0\simeq0.4$~eV, $a\simeq3\AA$.
One reason why the strong intensities are required 
derives from 
the fact that the time-reversal breaking terms evoked 
here are of fourth order in $E$, 
while usually the time-reversal breaking terms can be of second order~\cite{Claassen2017,Kitamura2017}.  
This comes from 
the cancellation similar to Fig.~\ref{fig:two-step}a 
 for the present square lattice [See Eq.~(\ref{eq:cancel}) for details], 
and can be evaded in e.g. honeycomb and kagome lattices. 
So the application of the present mechanism 
for the Floquet topological superconductivity to 
a wider class of materials with various 
lattice structures and/or multi-orbitals may 
be an interesting strategy.  
More trivially, going to low frequencies is another practical route for the enhanced $\gamma$, since the required intensity scales with the frequency $\omega$.  
A possibility of chiral superconductivity triggered by the same interaction but with a different mechanism
(such as proposed for the doped spin liquid on triangular lattice~\cite{Jiang2020}) is also of interest.

We can further raise an intriguing possibility that, since the emergence of the topological superconductivity is intimately related to the excitation of Higgs modes as stressed above, the topological signature 
might be enhanced by resonantly exciting the Higgs modes, 
which provides another future problem.

\appendix



\section*{Methods}

\subsection{Time-periodic Schrieffer-Wolff transformation}

To obtain the effective low-energy Hamiltonian in the presence of
the laser electric field, we employ the time-periodic Schrieffer-Wolff
transformation (a canonical transformation). Following a previous
study~\cite{Kitamura2017} (but extending it for 
the case where holes exist; see also Ref.~\cite{Kumar2021}), we decompose the Hubbard Hamiltonian as 
\begin{equation}
\hat{H}(t)=-\lambda\sum_{m=-\infty}^{\infty}(\hat{T}_{-1,m}+\hat{T}_{0,m}+\hat{T}_{+1,m})e^{-im\omega t}+U\hat{D},
\end{equation}
where $\lambda$ is a bookkeeping parameter bridging the atomic
limit $\lambda=0$ to the system of interest at $\lambda=1$, 
and 
$\hat{D}\equiv  \sum_{i}\hat{n}_{i\uparrow}\hat{n}_{i\downarrow}$ counts
the number of doubly-occupied sites. The hopping operator $\hat{T}_{d,m}$ 
describes the process that increases the number of doubly-occupied sites by $d$, as given by
\begin{align}
\hat{T}_{0,m} & =\sum_{ij\sigma}t_{ij}^{(m)}\left[\hat{n}_{i\bar{\sigma}}\hat{c}_{i\sigma}^{\dagger}\hat{c}_{j\sigma}\hat{n}_{j\bar{\sigma}}+(1-\hat{n}_{i\bar{\sigma}})\hat{c}_{i\sigma}^{\dagger}\hat{c}_{j\sigma}(1-\hat{n}_{j\bar{\sigma}})\right],\\
\hat{T}_{+1,m} & =\sum_{ij\sigma}t_{ij}^{(m)}\hat{n}_{i\bar{\sigma}}\hat{c}_{i\sigma}^{\dagger}\hat{c}_{j\sigma}(1-\hat{n}_{j\bar{\sigma}})=\hat{T}_{-1,-m}^{\dagger},\\
t_{ij}^{(m)} & =t_{ij}\frac{\omega}{2\pi}\int_{0}^{2\pi/\omega}dte^{-i\bm{A}(t)\cdot\bm{R}_{ij}+im\omega t} \\
 & =t_{ij}\mathcal{J}_{m}(A_{ij})(-1)^me^{im\theta_{ij}},\label{eq:hopping-fourier}
\end{align}
where $\bar{\sigma}\equiv -\sigma$.   
We also decompose the generator $\hat{\Lambda}(t)$ in Eq.~(\ref{eq:canonical})
into $e^{i\hat{\Lambda}(t)}=e^{i\hat{\Lambda}_{\text{h}}(t)}e^{i\hat{\Lambda}_{\text{c}}(t)}$,
where $\hat{\Lambda}_{\text{c}}$ eliminates the charge excitations (doubly-occupied sites) from the transformed Hamiltonian, 
while $\hat{\Lambda}_{\text{h}}$ makes the Hamiltonian static. 

We first consider $\hat{H}_{\text{c}}(t):=\hat{P}_{G}e^{i\hat{\Lambda}_{\text{c}}(t)}(\hat{H}(t)-i\partial_{t})e^{-i\hat{\Lambda}_{\text{c}}(t)}\hat{P}_{G}$.
Let us introduce a series solution,
\begin{equation}
\hat{\Lambda}_{\text{c}}(t)=\sum_{n=1}^{\infty}\sum_{d\neq0}\sum_{m=-\infty}^{\infty}\lambda^{n}\hat{\Lambda}_{+d,m}^{(n)}e^{-im\omega t}
\end{equation}
with $[\hat{D},\hat{\Lambda}_{+d,m}^{(n)}]=d\hat{\Lambda}_{+d,m}^{(n)}$,
which eliminates the charge excitation from $\hat{H}_{\text{c}}(t)$,
i.e., $[\hat{D},e^{i\hat{\Lambda}_{\text{c}}(t)}(\hat{H}(t)-i\partial_{t})e^{-i\hat{\Lambda}_{\text{c}}(t)}]=0$.
The form of $\hat{\Lambda}_{+d,m}^{(n)}$ can be uniquely determined
order by order, and we arrive at 
\begin{align}
\hat{H}_{\text{c}} & =-\lambda\sum_{m=-\infty}^{\infty}\hat{P}_{G}\hat{T}_{0,m}\hat{P}_{G}e^{-im\omega t}\nonumber \\
 & +\lambda^{2}\sum_{n,m=-\infty}^{\infty}\frac{\hat{P}_{G}\left[\hat{T}_{+1,n},\hat{T}_{-1,m-n}\right]\hat{P}_{G}}{2(U-n\omega)}e^{-im\omega t}+\text{H.c.}+O(\lambda^{3}).
\end{align}
While this Hamiltonian projected onto the spin subspace is evaluated
in a previous study\cite{Kitamura2017}, here we need to consider the expression for 
nonzero numbers of holes. We then obtain
\begin{align}
 & \hat{P}_{G}\left[\hat{T}_{+1,n},\hat{T}_{-1,m-n}\right]\hat{P}_{G}\nonumber \\
 & =-\sum_{ijk\sigma\sigma^{\prime}}\hat{P}_{G}(t_{ij}^{(m-n)}\hat{c}_{i\sigma}^{\dagger}\hat{c}_{j\sigma})\hat{n}_{j\bar{\sigma}}(t_{jk}^{(n)}\hat{c}_{j\sigma^{\prime}}^{\dagger}\hat{c}_{k\sigma^{\prime}})\hat{P}_{G}\\
 & =\sum_{ijk\sigma\sigma^{\prime}}t_{ij}^{(m-n)}t_{jk}^{(n)}\hat{P}_{G}\left[(\hat{c}_{i\sigma}^{\dagger}\bm{\sigma}_{\sigma\sigma^{\prime}}\hat{c}_{k\sigma^{\prime}})\cdot\hat{\bm{S}}_{j}-\frac{1}{2}\delta_{\sigma\sigma^{\prime}}\hat{c}_{i\sigma}^{\dagger}\hat{c}_{k\sigma}\hat{n}_{j}\right]\hat{P}_{G}.
\end{align}
Note that, while we recover the Heisenberg spin-spin interaction 
for $k=i$, 
the so-called \textit{three-site terms} with $k\neq i$ arise as well. The final
static Hamiltonian $\hat{H}_{\text{F}}=e^{i\hat{\Lambda}_{\text{h}}(t)}(\hat{H}_{\text{c}}-i\partial_{t})e^{-i\hat{\Lambda}_{\text{h}}(t)}$
is obtained by using the formula~\cite{Bukov2015,Mikami2016,Kitamura2017}
\begin{equation}
\hat{H}_{\text{F}}=\hat{H}_{\text{c},0}+\sum_{m\neq0}\frac{\left[\hat{H}_{\text{c},-m},\hat{H}_{\text{c},m}\right]}{2m\omega}+O(\omega^{-2}),
\end{equation}
where $\hat{H}_{\text{c},m} \equiv (\omega/2\pi)\int_{0}^{2\pi/\omega}dt\hat{H}_{\text{c}}e^{im\omega t}$.
The second term is calculated up to $\lambda^{2}$ as
\begin{align}
 & \sum_{m\neq0}\frac{\left[\hat{H}_{\text{c},-m},\hat{H}_{\text{c},m}\right]}{2m\omega}=\lambda^{2}\sum_{m\neq0}\frac{\hat{P}_{G}\left[\hat{T}_{0,-m},\hat{T}_{0,m}\right]\hat{P}_{G}}{2m\omega}\\
 & =\lambda^{2}\sum_{ijk\sigma\sigma^{\prime}}\sum_{m\neq0}\frac{t_{ij}^{(-m)}t_{jk}^{(m)}}{m\omega}\nonumber \\
 & \times\hat{P}_{G}\left[(\hat{c}_{i\sigma}^{\dagger}\bm{\sigma}_{\sigma\sigma^{\prime}}\hat{c}_{k\sigma^{\prime}})\cdot\hat{\bm{S}}_{j}+\delta_{\sigma\sigma^{\prime}}\hat{c}_{i\sigma}^{\dagger}\hat{c}_{k\sigma}\frac{2-\hat{n}_{j}}{2}\right]\hat{P}_{G},
\end{align}
 where we have used $\{\hat{c}_{i\sigma}(1-\hat{n}_{i\bar{\sigma}}),(1-\hat{n}_{j\bar{\sigma}^{\prime}})\hat{c}_{j\sigma^{\prime}}^{\dagger}\}=\delta_{ij}[\delta_{\sigma\sigma^{\prime}}(1-\hat{n}_{i}/2)+\bm{\sigma}_{\sigma\sigma^{\prime}}\cdot\hat{\bm{S}}_{i}]$,
$\{(1-\hat{n}_{i\bar{\sigma}})\hat{c}_{i\sigma}^{\dagger},(1-\hat{n}_{j\bar{\sigma}^{\prime}})\hat{c}_{j\sigma^{\prime}}^{\dagger}\}=0$. 

The effective Hamiltonian Eq.~(\ref{eq:canonical}) is
obtained up to the second order as 
\begin{align}
\hat{H}_{\text{F}} &
 =-\sum_{ij\sigma}\tilde{t}_{ij}\hat{P}_{G}\hat{c}_{i\sigma}^{\dagger}\hat{c}_{j\sigma}\hat{P}_{G}+\frac{1}{2}\sum_{ij}J_{ij}\left(\hat{\bm{S}}_{i}\cdot\hat{\bm{S}}_{j}-\frac{1}{4}\hat{n}_{i}\hat{n}_{j}\right)\nonumber \\
 & +\sum_{ijk\sigma\sigma^{\prime}}\Gamma_{i,j;\,k}\hat{P}_{G}\left[(\hat{c}_{i\sigma}^{\dagger}\bm{\sigma}_{\sigma\sigma^{\prime}}\hat{c}_{j\sigma^{\prime}})\cdot\hat{\bm{S}}_{k}-\frac{1}{2}\delta_{\sigma\sigma^{\prime}}\hat{c}_{i\sigma}^{\dagger}\hat{c}_{j\sigma}\hat{n}_{k}\right]\hat{P}_{G},
\end{align}
where 
\begin{align}
\tilde{t}_{ij} & =t_{ij}^{(0)}-\sum_{k}\sum_{m\neq0}\frac{t_{ik}^{(-m)}t_{kj}^{(m)}}{m\omega},\\
J_{ij} & =\sum_{m=-\infty}^{\infty}\frac{4U|t_{ij}^{(m)}|^{2}}{U^{2}-m^{2}\omega^{2}},\\
\Gamma_{i,j;\,k} & =\left[\frac{t_{ik}^{(0)}t_{kj}^{(0)}}{U}+\sum_{m\neq0}\frac{t_{ik}^{(-m)}t_{kj}^{(m)}}{m\omega(1-m\omega/U)}\right](1-\delta_{ij}).
\end{align}
Note that the second term in the hopping $\tilde{t}_{ij}$ describing the two-step hopping vanishes for the square lattice [See Fig.~\ref{fig:two-step}(a)], and the renormalised hopping is given as $\tilde{t}_{ij}=t_{ij}^{(0)}=t_{ij}\mathcal{J}_0(A_{ij})$ if we put $m=0$ in Eq.~(\ref{eq:hopping-fourier}). 
For the Heisenberg term $J_{ij}$ we end up with Eq.~(\ref{eq:heisenberg}) in the main text.
The two-step correlated hopping is much more intricate, 
and reads 
\begin{align}
\Gamma_{i,j;\,k} & =\frac{t_{ik}t_{kj}\mathcal{J}_{0}(A_{ik})\mathcal{J}_{0}(A_{kj})}{U}\nonumber\\
& +\sum_{m\neq0}\frac{t_{ik}t_{kj}\mathcal{J}_{m}(A_{ik})\mathcal{J}_{m}(A_{kj})}{m\omega(1-m\omega/U)}e^{im(\theta_{jk}-\theta_{ik})}.\label{eq:Gamma}
\end{align}

In the main text, we have also considered the scalar spin-chirality term $J_{ijk}^{\chi}$, which appears in the fourth-order perturbation.
While the perturbative processes involving three sites are considered in previous studies~\cite{Kitamura2017,Claassen2017},
we need to evaluate the processes involving four sites as well for the present case of the square lattice as we see in Eq.~(\ref{eq:cancel}) below.
By dropping the density dependent part, $\hat{\bm{S}}_i\cdot(\hat{\bm{S}}_j\times\hat{\bm{S}}_k)(1-\hat{n}_h)\sim O(\delta)$, 
we obtain
\begin{align}
J_{ijk}^{\chi} & =2\text{Im}\sum_{lmn=-\infty}^{\infty}(K_{ijk}^{lmn}+K_{jki}^{lmn}+K_{kij}^{lmn}-K_{jik}^{lmn}-K_{kji}^{lmn}-K_{ikj}^{lmn}),\label{eq:chirality}
\end{align}
where
\begin{align}
K_{ijk}^{lmn} & =\frac{2t_{ij}^{(-l-m)}t_{ji}^{(l-m)}t_{jk}^{(m-n)}t_{kj}^{(n+m)}(1-\delta_{m,0})}{m\omega[U+(l-m)\omega][U-(n+m)\omega]}\nonumber \\
 & -\frac{2t_{ij}^{(m-n)}t_{ji}^{(l)}t_{jk}^{(n)}t_{kj}^{(-l-m)}}{(U+l\omega)(U-m\omega)(U-n\omega)}\nonumber \\
 & -\sum_{h\neq ijk}\frac{t_{ij}^{(-l-m)}t_{jk}^{(l)}t_{kh}^{(m-n)}t_{hi}^{(n)}+t_{ij}^{(m-n)}t_{jk}^{(n)}t_{kh}^{(-l-m)}t_{hi}^{(l)}}{(U+l\omega)(U+n\omega)[U+(n-m)\omega]}\nonumber \\
 & +2\sum_{h\neq ijk}\frac{\left[(t_{ij}^{(l)}t_{jk}^{(-l-m)}+t_{ij}^{(-l-m)}t_{jk}^{(l)})t_{kh}^{(m-n)}+t_{ij}^{(m-n)}t_{jk}^{(l)}t_{kh}^{(-l-m)}\right]t_{hi}^{(n)}}{(U+l\omega)(U-n\omega)(U-m\omega)}\nonumber \\
 & +\sum_{h\neq ijk}\frac{t_{ij}^{(m-n)}t_{jk}^{(-l-m)}t_{kh}^{(l)}t_{hi}^{(n)}+t_{ij}^{(n)}t_{jk}^{(l)}t_{kh}^{(-l-m)}t_{hi}^{(m-n)}}{(U+l\omega)(U-n\omega)(U-m\omega)}.\label{eq:klmn}
\end{align}
The above expression is derived under an assumption $t_{ji}^{(n)}=t_{ij}^{(n)}(-1)^n$, which holds  for monochromatic laser lights. 
In particular, the chiral coupling of crucial interest 
reads, for the present square lattice with second-neighbour hopping,
\begin{align}
J_{i,i+y,i+x}^{\chi} & =J_\chi =2J_\chi^\prime+2J_\chi^{\prime\prime},\label{eq:chirality-details}\\
J_{i-x,i,i+x+y}^{\chi} & =J_\chi^\prime =4t_{0}^{2}t_{0}^{\prime2}\sum_{lmn}\sin\frac{m\pi}{2}\nonumber \\
 & \times\Biggl[\frac{D_{m-l}}{m\omega}+(D_{m-l}+2D_{n-l})D_{n-m}\Biggr]D_{n+m}\nonumber \\
 & \times\Biggl[\mathcal{J}_{l+m}(\sqrt{2}A)\mathcal{J}_{l-m}(\sqrt{2}A)\mathcal{J}_{n+m}(A)\mathcal{J}_{n-m}(A)\nonumber \\
 & +\mathcal{J}_{n+m}(\sqrt{2}A)\mathcal{J}_{n-m}(\sqrt{2}A)\mathcal{J}_{l+m}(A)\mathcal{J}_{l-m}(A)\Biggr],\\
J_{i-x,i-y,i+x}^{\chi} & =J_\chi^{\prime\prime} =4t_{0}^{2}t_{0}^{\prime2}\sum_{lmn}\sin\frac{(3m+2n)\pi}{4}\nonumber \\
 & \times\Biggl[D_{l}D_{n}D_{n-m}+(D_{l}D_{-n}+D_{-n}D_{-n+m})D_{l+m-n}\Biggr]\nonumber \\
 & \times\Biggl[\mathcal{J}_{-l-m}(A)\mathcal{J}_{l}(A)\mathcal{J}_{m-n}(\sqrt{2}A)\mathcal{J}_{n}(\sqrt{2}A)\nonumber \\
 & +\mathcal{J}_{m-n}(A)\mathcal{J}_{n}(A)\mathcal{J}_{-l-m}(\sqrt{2}A)\mathcal{J}_{l}(\sqrt{2}A)\Biggr],
\end{align}
where $D_{n}=(U+n\omega)^{-1}$ and $A=Ea/\omega$ with $a$ being the lattice constant.

In Discussion we mentioned that 
the time-reversal breaking terms are of fourth order in $E$ 
due to a cancellation similar to Fig.~\ref{fig:two-step}a.  
Let us here elaborate on that, which marks the importance of the perturbative processes involving four sites. If we expand Eq.~(\ref{eq:klmn}) up to $E^2$, 
we can find
\begin{align}
&\text{Im}\sum_{lmn=-\infty}^{\infty}(K_{ijk}^{lmn}-K_{kji}^{lmn})\nonumber\\
&\propto i(\bm{E}^{\ast}\times\bm{E})\cdot\left(t_{kj}t_{ji}\bm{R}_{ij}\times\bm{R}_{jk}+\sum_{h\neq ijk}t_{kh}t_{hi}\bm{R}_{ih}\times\bm{R}_{hk}\right)t_{ij}t_{jk}.\label{eq:cancel}
\end{align}
The first term on the second line (three-site term) coincides with the result in a previous study~\cite{Kitamura2017}.
While the second term (four-site term) is absent for e.g. honeycomb and kagome lattices with no candidates for the fourth site $h$, 
square lattice accommodates this additional contribution. 
Then we can note that 
we have a combination of blue and red paths as 
 depicted in Fig.~\ref{fig:two-step}a, 
which was invoked for the three-site term but 
also applies to the four-site term involving 
$t_{kj}t_{ji}$ and $t_{kh}t_{hi}$.  There, due to 
the factor $\bm{R}\times\bm{R}$, the first and second terms cancel with each other for the square lattice (even if we consider many-body effects).  
The square lattice model does have a nonzero coefficient however, 
if we go over to $E^4$ in the presence of the next-nearest-neighbour hopping, 
as we have seen Eq.~(\ref{eq:Jchi-fourth}).

\subsection{Gutzwiller projection}

Here we analyse the ground-state property of the system using the mean-field approximation with the Gutzwiller ansatz~\cite{Ogata2003,Ogata2008}.
Namely, we consider an ansatz for the wavefunction,
\begin{align}
|\Psi\rangle & =\hat{P}_G|\Psi_0\rangle,\\
|\Psi_0\rangle & =\prod_{\bm{k}}\left(u_{\bm{k}}+v_{\bm{k}}\hat{c}_{\bm{k}\uparrow}^\dagger\hat{c}_{-\bm{k}\downarrow}^\dagger\right)|0\rangle,\label{eq:BCS}
\end{align}
with $\hat{P}_G=\prod_i(1-\hat{n}_{i\uparrow}\hat{n}_{i\downarrow})$ being the Gutzwiller projection. Here, $\hat{c}_{\bm{k}\sigma}=N^{-1/2}\sum_i \hat{c}_{i\sigma} e^{-i\bm{k}\cdot\bm{R}_i}$ with $N$ being the number of lattice sites.
The ground state within this ansatz can be obtained by minimising the expectation value, 
\begin{align}
E_0=\frac{\langle\Psi|\hat{H}_{\text{F}}|\Psi\rangle}{\langle\Psi|\Psi\rangle}=\frac{\langle\hat{P}_G\hat{H}_{\text{F}}\hat{P}_G\rangle_0}{\langle\hat{P}_G\rangle_0}.
\end{align}

To evaluate the Gutzwiller projection approximately, we replace
the expectation value with the site-diagonal ones, where we decompose the projection operator as 
\begin{align}
\hat{P}_{G} & =\sum_{c_{1},\dots,c_{N}=h,\uparrow,\downarrow}\left(\prod_{i=1}^N\hat{P}_{ic_{i}}\right)
\end{align}
with $\hat{P}_{ih}\equiv(1-\hat{n}_{i\uparrow})(1-\hat{n}_{i\downarrow}),\hat{P}_{i\sigma}\equiv n_{i\sigma}(1-n_{i\bar{\sigma}})$.
Then we can evaluate the denominator as
\begin{align}
\langle \hat{P}_{G}\rangle_{0} & \simeq \sum_{c_{1},\dots,c_{N}=h,\uparrow,\downarrow}\prod_{i=1}^N\langle \hat{P}_{ic_{i}}\rangle_{0}\\
 & =\frac{N!}{(N\delta)!(Nf)!(Nf)!}(\bar{f}\bar{f})^{N\delta}(f\bar{f})^{Nf}(f\bar{f})^{Nf}\\
 & \sim \frac{(\bar{f}\bar{f})^{N\delta}(f\bar{f})^{Nf}(f\bar{f})^{Nf}}{\delta^{N\delta}f^{Nf}f^{Nf}}=(\bar{f}^2\delta^{-1})^{N\delta}\bar{f}^{2Nf},
\end{align}
where $f=(1-\delta)/2,\bar{f}=(1+\delta)/2$, and we have used $n!\sim (n/e)^n$ on the last line.
We can evaluate the numerators in the same manner. For the hopping, exchange, and spin-dependent three-site terms, we obtain, respectively,
\begin{align}
\langle \hat{P}_{G}\hat{c}_{i\sigma}^{\dagger}\hat{c}_{j\sigma}\hat{P}_{G}\rangle_{0} & 
\simeq(\bar{f}^2\delta^{-1})^{N\delta-1}\bar{f}^{2Nf-1}\langle\bar{f}\hat{c}_{i\sigma}^{\dagger}\hat{c}_{j\sigma}\bar{f}\rangle_{0}\\
 & =\frac{2\delta}{1+\delta}\langle \hat{c}_{i\sigma}^{\dagger}\hat{c}_{j\sigma}\rangle_{0}\langle \hat{P}_{G}\rangle_{0},\\
\langle \hat{P}_{G}\hat{\bm{S}}_{i}\cdot\hat{\bm{S}}_{j}\hat{P}_{G}\rangle_{0} & 
\simeq(\bar{f}^2\delta^{-1})^{N\delta}\bar{f}^{2Nf-2}\langle\hat{\bm{S}}_{i}\cdot\hat{\bm{S}}_{j}\rangle_{0}\\
 & =\frac{4}{(1+\delta)^{2}}\langle\hat{\bm{S}}_{i}\cdot\hat{\bm{S}}_{j}\rangle_{0}\langle \hat{P}_{G}\rangle_{0},\\
\langle\hat{P}_{G}\hat{c}_{i\sigma}^{\dagger}\hat{c}_{j\sigma^{\prime}}\hat{\bm{S}}_{k}\hat{P}_{G}\rangle_{0} &
 \simeq(\bar{f}^2\delta^{-1})^{N\delta-1}\bar{f}^{2Nf-2}\langle\bar{f}\hat{c}_{i\sigma}^{\dagger}\hat{c}_{j\sigma^{\prime}}\bar{f}\hat{\bm{S}}_{k}\rangle_{0}\\
 & =\frac{4\delta}{(1+\delta)^{2}}\langle\hat{c}_{i\sigma}^{\dagger}\hat{c}_{j\sigma^{\prime}}\hat{\bm{S}}_{k}\rangle_{0}\langle \hat{P}_{G}\rangle_{0}.
\end{align}
We also obtain $\langle \hat{P}_{G}\hat{\bm{S}}_{i}\cdot(\hat{\bm{S}}_{j}\times\hat{\bm{S}}_{k})\hat{P}_{G}\rangle_{0}  
 \simeq [8/(1+\delta)^{3}]\langle\hat{\bm{S}}_{i}\cdot(\hat{\bm{S}}_{j}\times\hat{\bm{S}}_{k})\rangle_{0}\langle \hat{P}_{G}\rangle_{0}
$ in the same manner.

The remaining terms have an ambiguity in the Gutzwiller factor,
because they are accompanied by the density operator $\hat{P}_G\hat{n}_i\hat{P}_G=\hat{P}_G[\sum_{\sigma}\hat{n}_{i\sigma}(1-\hat{n}_{i\bar{\sigma}})]\hat{P}_G$, 
which can be regarded either as an operator or a constant $\langle\hat{n}_{i\sigma}(1-\hat{n}_{i\bar{\sigma}})\rangle_0\simeq f\bar{f}$, in the above scheme.
Here, we discard the second-order fluctuation around the expectation value, $(\hat{n}_{i\sigma}-f)(1-\hat{n}_{i\bar{\sigma}}-\bar{f})$, to approximate the projected density operator as $\sum_{\sigma}\hat{n}_{i\sigma}(1-\hat{n}_{i\bar{\sigma}})\simeq\hat{n}_{i}\delta+2f^2$. 
Then the Gutzwiller factors for the remaining terms are evaluated as
\begin{multline}
\langle \hat{P}_{G}\hat{n}_{i}\hat{n}_{j}\hat{P}_{G}\rangle_{0}  \simeq\frac{4}{(1+\delta)^{2}}\langle(\hat{n}_{i}\delta+2f^{2})(\hat{n}_{j}\delta+2f^{2})\rangle_{0}\langle \hat{P}_{G}\rangle_{0}\\
  =\frac{4\delta^{2}}{(1+\delta)^{2}}\langle\hat{n}_{i}\hat{n}_{j}\rangle_{0}\langle \hat{P}_{G}\rangle_{0}
  +\frac{2\delta(1-\delta)^{2}}{(1+\delta)^{2}}\langle\hat{n}_{i}+\hat{n}_{j}\rangle_{0}\langle \hat{P}_{G}\rangle_{0}+\text{const.},
\end{multline}
\begin{multline}
\langle\hat{P}_{G}\hat{c}_{i\sigma}^{\dagger}\hat{c}_{j\sigma}\hat{n}_{k}\hat{P}_{G}\rangle_{0}  \simeq\frac{4\delta}{(1+\delta)^{2}}\langle\hat{c}_{i\sigma}^{\dagger}\hat{c}_{j\sigma}(\hat{n}_{k}\delta+2f^{2})\rangle_{0}\langle \hat{P}_{G}\rangle_{0}\\
  =\frac{4\delta^{2}}{(1+\delta)^{2}}\langle\hat{c}_{i\sigma}^{\dagger}\hat{c}_{j\sigma}\hat{n}_{k}\rangle_{0}\langle \hat{P}_{G}\rangle_{0} +\frac{2\delta(1-\delta)^{2}}{(1+\delta)^{2}}\langle\hat{c}_{i\sigma}^{\dagger}\hat{c}_{j\sigma}\rangle_{0}\langle \hat{P}_{G}\rangle_{0}.
\end{multline}
Note that the density-density term is known to have small effects in the variational Monte Carlo calculation in the low-doping regime, 
with which the present treatment is consistent.

With these expressions, the minimisation of $E_0$ turns out to be reduced to that of $\langle\hat{H}_{\text{F}0}\rangle_0$, where $\hat{H}_{\text{F}0}$ is the Hamiltonian with the modified coupling constant but without the Gutzwiller projection, as given by
\begin{align}
\hat{H}_{\text{F}0} & =-g\delta\sum_{ij\sigma}\left[\tilde{t}_{ij}+\frac{(1-\delta)^{2}}{2(1+\delta)}\sum_{k}\Gamma_{ijk}\right]\hat{c}_{i\sigma}^{\dagger}\hat{c}_{j\sigma}\nonumber \\ 
 & +\frac{g^2}{2}\sum_{ij}J_{ij}\left(\hat{\bm{S}}_{i}\cdot\hat{\bm{S}}_{j}-\frac{\delta^{2}}{4}\hat{n}_{i}\hat{n}_{j}\right)
 +\frac{g^3}{6}\sum_{ijk}J^\chi_{ijk}(\hat{\bm{S}}_{i}\times\hat{\bm{S}}_{j})\cdot\hat{\bm{S}}_{k}\nonumber \\
 & +g^2\delta\sum_{ijk\sigma\sigma^{\prime}} \Gamma_{ijk}\left[(\hat{c}_{i\sigma}^{\dagger}\bm{\sigma}_{\sigma\sigma^{\prime}}\hat{c}_{j\sigma^{\prime}})\cdot\hat{\bm{S}}_{k}-\frac{\delta}{2}\delta_{\sigma\sigma^{\prime}}\hat{c}_{i\sigma}^{\dagger}\hat{c}_{j\sigma}\hat{n}_{k}\right]
\end{align}
with 
\begin{align}
g=\frac{2}{1+\delta}.
\end{align}
Differenciating $\langle\hat{H}_{\text{F}0}\rangle_0$ by $u_{\bm{k}}, v_{\bm{k}}$ in Eq.~(\ref{eq:BCS}),
we arrive at the Bogoliubov-de Gennes Hamiltonian Eq.~(\ref{eq:BdG}) in the main text.
In particular, the detailed form for the present system is 
given as
\begin{align}
\varepsilon(\bm{k}) & =-\mu+\frac{1}{2}\varepsilon_{x}(\cos k_{x}+\cos k_{y})+\varepsilon_{x+y}\cos k_{x}\cos k_{y}\nonumber \\
 & -4g\delta(1-\delta)\text{Re}\Gamma_{i-x-y,i+x;i}(\cos2k_{x}\cos k_{y}+\cos k_{x}\cos2k_{y})\nonumber \\
 & -g\delta(1-\delta)\text{Re}(2\Gamma_{i-x,i+x;i+y}+\Gamma_{i-x,i+x;i})(\cos2k_{x}+\cos2k_{y})\nonumber \\
 & -2g\delta(1-\delta)\text{Re}\Gamma_{i-x-y,i+x+y;i}\cos2k_{x}\cos2k_{y}\nonumber \\
 & +\frac{3}{2}g^{3}J_{\chi}^{\prime}\text{Re}(\Delta^{\ast}\Delta^{\prime})(\cos2k_{x}\cos k_{y}+\cos k_{x}\cos2k_{y}),\\
\varepsilon_{x} & =-4g\delta\tilde{t}-\frac{3-\delta^{2}}{2}g^{2}J\chi_{x}-8g\delta(1-\delta)\text{Re}\Gamma_{i-x,i;i+y}\nonumber \\
 & -2g^{2}\delta(3-\delta)\text{Re}(2\Gamma_{i-x,i+y;i}+\Gamma_{i-x,i+x;i})\chi_{x}\nonumber \\
 & -4g^{2}\delta(3-\delta)\text{Re}(\Gamma_{i-x,i;i+y}+\Gamma_{i-x-y,i+x;i})\chi_{x+y}\nonumber \\
 & +\frac{3}{2}g^{3}\text{Re}\left[2J_{\chi}\Delta^{\prime}\Delta^{\ast}+J_{\chi}^{\prime}\Delta^{\prime}(\Delta_{x+2y}^{\ast}-\Delta_{2x+y}^{\ast})\right],\\
\varepsilon_{x+y} & =-4g\delta\tilde{t}^{\prime}-\frac{3-\delta^{2}}{2}g^{2}J^{\prime}\chi_{x+y}-4g\delta(1-\delta)\text{Re}\Gamma_{i-x,i+y;i}\nonumber \\
 & -2g^{2}\delta(3-\delta)\text{Re}(2\Gamma_{i-x,i+x;i+y}+\Gamma_{i-x-y,i+x+y;i})\chi_{x+y}\nonumber \\
 & -4g^{2}\delta(3-\delta)\text{Re}(\Gamma_{i-x,i;i+y}+\Gamma_{i-x-y,i+x;i})\chi_{x}\nonumber \\
 & -\frac{3}{2}g^{3}\text{Re}\left[2J_{\chi}^{\prime\prime}\Delta_{2x}^{\ast}\Delta^{\prime}-iJ_{\chi}^{\prime}\Delta^{\ast}(\Delta_{2x+y}+\Delta_{x+2y})\right],
\end{align}
\begin{align}
F(\bm{k}) & =\frac{1}{2}F_{x}(\cos k_{x}-\cos k_{y})+iF_{x+y}\sin k_{x}\sin k_{y}\nonumber \\
 & +\frac{3g^{3}J_{\chi}^{\prime\prime}}{2}\Delta^{\prime}\chi_{x+y}(\cos2k_{x}-\cos2k_{y})\nonumber \\
 & -\frac{3ig^{3}J_{\chi}^{\prime}}{2}\Delta\chi_{x+y}(\sin2k_{x}\sin k_{y}+\sin k_{x}\sin2k_{y})\nonumber \\
 & -\frac{3g^{3}J_{\chi}^{\prime}}{2}\chi_{x}\Delta^{\prime}(\cos k_{x}\cos2k_{y}-\cos2k_{x}\cos k_{y}),\label{eq:gap-full}\\
F_{x} & =g^{2}\left[\frac{3+\delta^{2}}{2}J-2\delta(3+\delta)\text{Re}(2\Gamma_{i-x,i+y;i}-\Gamma_{i-x,i+x;i})\right]\Delta\nonumber \\
 & -g^{2}\left[4\delta(3+\delta)\gamma+3gJ_{\chi}\chi_{x}+\frac{3}{2}gJ_{\chi}^{\prime}(\chi_{2x+y}+\chi_{x+2y})\right]\Delta^{\prime}\nonumber \\
 & -\frac{3}{2}ig^{3}J_{\chi}^{\prime}(\Delta_{2x+y}+\Delta_{x+2y})\chi_{x+y},\\
F_{x+y} & =g^{2}\left[4\delta(3+\delta)\gamma+3gJ_{\chi}\chi_{x}+\frac{3}{2}gJ_{\chi}^{\prime}(\chi_{2x+y}+\chi_{x+2y})\right]\Delta\nonumber \\
 & -g^{2}\left[\frac{3+\delta^{2}}{2}J^{\prime}-2\delta(3+\delta)\text{Re}(2\Gamma_{i-x,i+x;i+y}-\Gamma_{i-x-y,i+x+y;i})\right]\Delta^{\prime} \nonumber \\
 & -\frac{3}{2}g^{3}\left[J_{\chi}^{\prime}\chi_{x}(\Delta_{2x+y}-\Delta_{x+2y})+2J_{\chi}^{\prime\prime}\Delta_{2x}\chi_{x+y}\right].
\end{align}
When $\Delta$ is real, we can further simplify the expressions using $\chi_{mx+ny}=\chi_{nx+my}$ and $\Delta_{mx+ny}=-\Delta_{nx+my}^\ast$,
after which the first line of $F_{x+y}$ gives $F^{xy}(\bm{k})$ in the main text.

\begin{acknowledgments}
S.K. acknowledges JSPS KAKENHI Grant  20K14407, 
and CREST (Core Research for Evolutional
Science and Technology; Grant number JPMJCR19T3) for support.  
H.A. thanks CREST (Grant Number JPMJCR18T4), and JSPS KAKENHI Grant JP17H06138.
\end{acknowledgments}

\end{document}